\documentclass[12pt]{article}
\usepackage{hyperref}

\usepackage{varioref}

\IfFileExists{srcltx.sty}{\usepackage[active]{srcltx}}

\catcode`\@=11
\def\marginnote#1{}

\newcount\hour
\newcount\minute
\newtoks\amorpm
\hour=\time\divide\hour by60
\minute=\time{\multiply\hour by60 \global\advance\minute by-\hour}
\edef\standardtime{{\ifnum\hour<12 \global\amorpm={am}%
           \else\global\amorpm={pm}\advance\hour by-12 \fi
           \ifnum\hour=0 \hour=12 \fi
           \number\hour:\ifnum\minute<10 0\fi\number\minute\the\amorpm}}
\edef\militarytime{\number\hour:\ifnum\minute<10 0\fi\number\minute}

\def\draftlabel#1{{\@bsphack\if@filesw {\let\thepage\relax
      \xdef\@gtempa{\write\@auxout{\string
         \newlabel{#1}{{\@currentlabel}{\thepage}}}}}\@gtempa
      \if@nobreak \ifvmode\nobreak\fi\fi\fi\@esphack}
           \gdef\@eqnlabel{#1}}
\def\@eqnlabel{}
\def\@vacuum{}
\def\draftmarginnote#1{\marginpar{\raggedright\tiny\tt#1}}

\def\draftlabel#1{{\@bsphack\if@filesw {\let\thepage\relax
      \xdef\@gtempa{\write\@auxout{\string
         \newlabel{#1}{{\@currentlabel}{\thepage}}}}}\@gtempa
      \if@nobreak \ifvmode\nobreak\fi\fi\fi\@esphack}
           \gdef\@eqnlabel{#1}}
\def\@eqnlabel{}
\def\@vacuum{}
\def\draftmarginnote#1{\marginpar{\raggedright\scriptsize\tt #1}}

\def\draft{\oddsidemargin -.5truein
           \def\@oddfoot{{\footnotesize Draft of {\tt\jobname.tex}}
             \hfil
           -- \thepage\ -- \hfil{\footnotesize\today\quad\militarytime}}
           \let\@evenfoot\@oddfoot \overfullrule 3pt
           \let\label=\draftlabel
           \let\marginnote=\draftmarginnote
      \def\@eqnnum{(\theequation)\rlap{\kern\marginparsep\footnotesize\tt\@eqnlabel}%
\global\let\@eqnlabel\@vacuum}  }

\def\numberbysection{\@addtoreset{equation}{section}
           \def\theequation{\thesection.\arabic{equation}}}

\def\underline#1{\relax\ifmmode\@@underline#1\else
           $\@@underline{\hbox{#1}}$\relax\fi}

\def\titlepage{\@restonecolfalse\if@twocolumn\@restonecoltrue\onecolumn
        \else \newpage \fi \thispagestyle{empty}\c@page\z@
           \def\thefootnote{\fnsymbol{footnote}} }

\def\endtitlepage{\if@restonecol\twocolumn \else  \fi
           \def\thefootnote{\arabic{footnote}}
           \setcounter{footnote}{0}}  
\relax

\usepackage{amsfonts}
\newcommand{\IC}{\mathbb{C}} 
\newcommand{\IN}{\mathbb{N}}

\def\h{\hbar}
\newcommand{\gl}[1]{\ensuremath{GL(#1,\IC)}}

\newcommand{\vt}{\textbf{{\textit{t}}}}

\newcommand{\la}{\langle}
\newcommand{\ra}{\rangle}
\newcommand{\ket}[1]{\left|{#1}\right\rangle}
\newcommand{\bra}[1]{\left\langle{#1}\right|}

\newcommand{\dfrac}[2]{\displaystyle\frac{#1}{#2}}
\newcommand{\pfrac}[2]{\displaystyle\frac{\p #1}{\p #2}}
\newcommand{\p}{\partial}

\newcommand{\bz}[1]{\bar z_{#1}}

\newcommand{\ha}[1] {\ensuremath{a_{#1}}}
\newcommand{\ba}[1]{\ensuremath{{ \bar a}_{#1}}}

\newcommand{\at}[1] {\ensuremath{{\bar a}_{#1}}}
\newcommand{\hb}[1]{\ensuremath{ b_{#1}}}
\newcommand{\hbt}[1]{\ensuremath{{ \bar b}_{#1}}}
\newcommand{\bb}[1]{\ensuremath{{ \bar b}_{#1}}}

\renewcommand{\t}[1]{t_{#1}}
\newcommand{\bt}[1]{\bar t_{#1}}
\newcommand{\tb}[1]{\bar t_{#1}}

\def\CC{\ensuremath{{\cal C}}}

\def\CH{{\cal H}}
\def\CJ{{\cal J}}

\def\CO{{\cal O}}

\def\bea{\begin{eqnarray}}
\def\eea{\end{eqnarray}}

\def \de{\partial}
\def \p{\partial}

\def \bz {{\bar z}}
\def \bw {{\bar w}}
\def \bN {{\bar N}}

\def \bJ {{\bar J}}

\def\LB{\left(}
\def\RB{\right)}
\def\be{\begin{equation}}
\def\ee{\end{equation}}

\newcommand{\bket}[1]{\left|{#1}\rangle\rangle\right.}
\newcommand{\bbra}[1]{\left.\langle\langle{#1}\right|}
\newcommand{\sprod}[2]{\langle{#1}| {#2}\rangle}

\catcode`\@=11
\def\numberbysection{\@addtoreset{equation}{section}
           \def\theequation{\thesection.\arabic{equation}}}
\catcode`\@=12

\numberbysection

\begin{document} 

\title{\hfill\vbox{\normalsize\hbox{hep-th/0307057}    
    \hbox{AEI-2003-042}
    \hbox{EFI-03-20}%
    }\\
  \Large  String Field Theory Vertices, Integrability\\ and Boundary States} 
\author{ Alexey Boyarsky\footnote{boyarsky@alf.nbi.dk}~\footnote{On leave of
    absence from Bogolyubov ITP, Kiev, Ukraine}\\                        
  { \normalsize \it Niels Bohr Institute} \\
  {\normalsize \it Blegdamsvej 17, DK-2100 Copenhagen, Denmark}\\\\
  Bogdan Kulik\footnote{bogdansk@aei.mpg.de} \\
    {\normalsize \it Max-Planck-Institut f\"ur Gravitationsphysik}\\
    {\normalsize \it Albert-Einstein-Institut}\\
    {\normalsize \it Am M\"uhlenberg 1, D-14476 Golm, Germany}\\\\
  Oleg Ruchayskiy\footnote{ruchay@flash.uchicago.edu} \\
    {\normalsize \it Enrico Fermi Institute and Dept. of Physics}\\
    {\normalsize \it University of Chicago} \\
    {\normalsize \it 5640 Ellis Ave. Chicago IL, 60637, U.S.A.}}
\date{\small\today}
\maketitle

\begin{abstract}
  We study Neumann coefficients of the various vertices in the Witten's open
  string field theory (SFT). We show that they are not independent, but
  satisfy an infinite set of algebraic relations. These relations are
  identified as so-called Hirota identities. Therefore, Neumann coefficients
  are equal to the second derivatives of tau-function of dispersionless Toda
  Lattice hierarchy (this tau-function is just a partition sum of
  normal matrix model). As a result, certain two-vertices of SFT are
  identified with the Neumann boundary states on an arbitrary curve.

  We further analyze a class of SFT surface states, which can be re-written in
  the closed string language in terms of boundary states.  This offers a new
  correspondence between open string states and closed string states (boundary
  states) in SFT.  We conjecture that these special states can be considered
  as describing D-branes and other extended objects as "solitons" in SFT. We
  consider some explicit examples, one of them is a surface states
  corresponding to orientifold.
\end{abstract}

\tableofcontents

\section{Introduction and summary of main results}

A new structure seems to be underlying the open string field theory
(OSFT)~\cite{wittenSFT}. Namely, it becomes increasingly clear that a number
of objects of SFT can be given an interpretation in terms of \emph{integrable
  hierarchies}.

On the other hand, studies of tachyon dynamics and other related hypotheses
(see e.g.~\cite{sen}) within the scope of open SFT (see
e.g.~\cite{SenBr,vsft}) have risen the question of description of closed
string states in the open SFT.  Surprisingly, traveling an unexploited path of
integrability in SFT, we in fact will be able to address these questions as
well.

It has been shown in~\cite{Bo-Ru} that various states in CFT can be associated
with tau-functions of dispersionless KP and Toda Lattice
hierarchies\footnote{%
  Throughout this paper we will often use terms ``dKP'' and ``dToda'' when
  referring to the dispersionless KP and the dispersionless
  Toda Lattice hierarchies~\cite{tak-tak}  correspondingly.}.
Times of integrable hierarchy parameterize the dependence of these states on
an arbitrary conformal transformation.  Vacuum state for the holomorphic
scalar field in the plane (``open string picture'') corresponds to dKP
hierarchy\footnote{%
  \label{fn:7}Up to some subtleties, dKP hierarchy can
  be thought of as a holomorphic sector of dToda hierarchy.}.
Boundary states in case of arbitrary scalar field in the plane (``closed
string picture'') are described by the whole dToda hierarchy, with holomorphic
and antiholomorphic sectors being mixed. 

An important example of this construction  is a surface state
in open SFT~\cite{peskin}. Neumann coefficients of the
surface state were shown in~\cite{Bo-Ru} to be not independent but to 
satisfy an infinite set of
algebraic relations. These relations are nothing else but so-called
\emph{Hirota identities} for dKP hierarchy. It means that the Neumann
coefficients are just second derivatives of (dispersionless) KP tau-function.

Hirota identities distinguish the tau-functions of dispersionless integrable
hierarchies from any other function of infinite number of variables. 
They turned out to be equivalent to the condition for a state to be Bogolyubov
transform of a vacuum in case of ``open string picture''.  Namely, on the CFT
side Hirota identities are the conditions which guarantee existence of some
operators $b_k, \bar b_k$ annihilating the transformed state.  These operators
are linear combinations of original creation and annihilation operators $a_k,
\bar a_k$.

For the ``closed string picture'' the Hirota identities of dispersionless Toda
Lattice hierarchy mean that corresponding state is annihilated by combinations
$b_k \pm \bar b_k$, being thus Bogolyubov transform of Neumann or Dirichlet
boundary state.  Such a state is exponential of combination quadratic in
creation operators $a_k^+, \bar a_k^+$ with coefficients being second
derivatives of dispersionless Toda Lattice tau-function with respect to all
times $t_k$ and $\bar t_k$. It is worth noting that
this is the same tau-function that is equal to a partition sum of
normal matrix model \cite{Adler,NMM}.

Later, it was observed in~\cite{Bonora} that Neumann coefficients of Witten's
three-vertex in open SFT could also be expressed via derivatives of particular
tau-function of dispersionless Toda Lattice hierarchy.  Thus boundary states
in ``closed string picture'' and three-vertex in open SFT are expressed in
terms of the same data and should be somehow related.  However, naively there
is no immediate connection between these two objects.

Moreover, a three-vertex in SFT naturally depends on three independent sets of
creation operators~\cite{peskin}. This gives nine infinite matrices of the
corresponding Neumann coefficients $N^{IJ}$ ($I,J=1,2,3$).  On the other hand
in dToda hierarchy there are only two infinite sets of times $t_k, \bar t_k$
and, consequently, three infinite matrices of second derivatives
$F^{AB}\equiv\p_{t^{(A)}}\p_{t^{(B)}}F$, where $A,B=1,2$ and correspond to
$t_k$ and $\bar t_k$.  This difficulty did not appear in~\cite{Bonora} because
for the particular case of Witten's vertex (corresponding to special choice of
conformal maps which define gluing of three open string world sheets) all
matrices of the Neumann coefficients are expressed via just two independent
ones.  Still, even for this case the three-vertex depends on three independent
sets of creation operators, although some coefficients in front of them
coincide.  Because of this, the exact identification between combinations of
Neumann coefficients and second derivatives of tau-functions were just guessed
so that they would satisfy Hirota identities. The meaning of these formulae
was unclear. Also, it was unclear if there was any relation between the
three-vertex and boundary states.

In this paper we address these questions.  We find explicit connection between
the SFT two-vertices and boundary states.  By its nature a two-vertex relates
correlators in the tensor product of two open strings to correlators of one
open string.  On the other hand, a boundary state relates closed strings to
open strings. In order for a two-vertex to be a boundary state it should
combine two open strings into one closed. To achieve this, conformal
transformations that define the two-vertex should map world sheets of two open
strings into that of a closed string.  We state a simple condition such
transformations should obey.  Using this connection, we derive in a systematic
way the formulae relating Neumann coefficients for general SFT vertex and
second derivatives of tau-function. This relation is a trivial consequence of
the fact, that any $n$-vertex, contracted with vacua states in $n-2$ sectors
is just a two-vertex in the remaining two sectors.  In particular, this
explains the results guessed in~\cite{Bonora} for the case of Witten's
three-vertex.

  There is one more conceptual problem in~\cite{Bonora} which we would like
  to address here.  In the case of Witten's three-vertex that we considered
  till now all three conformal maps were fixed. It means that corresponding
  Neumann coefficients are just numbers, not functions of $t_k$, which are
  identified with second derivatives of tau function evaluated at some
  particular fixed values of $t_k$.  However, the whole experience of
  integrable systems (in its application e.g. to matrix models, SYM theories,
  quantum Hall effect) teaches us that it is important and usually fruitful to
  introduce the dynamics w.r.t. these $t_k$ even if we are interested at the
  end only in the results for some fixed times. (We will comment on the
  application of this ideology to the present case latter).  This is why, once
  we have found a tau function (or its second derivatives) calculated at some
  particular values, it is interesting to try to understand what could this
  object mean at different, arbitrary values of $t_k$.  Note, that for the
  ``chiral'' case of KP\footnote{See footnote~\vref{fn:7}} we have already
  found an answer to this question.  KP case was related to the surface states
  (or ``conformally transformed vacuum''~\cite{Bo-Ru}).  These states
  (one-vertices) $\ket{\Sigma}$ are defined for an arbitrary conformal map
  from some proper family.  We can parameterize this family of states
  $\ket{\Sigma(t_k)}$ by times $t_k$ and in such way describe dynamics of
  surface states. Thus the functional dependence of KP tau-function with
  respect to $t_k$ (not just its value at fixed $t_k$) makes an appearance in
  SFT.\@ We will discuss the consequences of this below.
  
  However, we are interested in finding the full Toda dynamics in the
  SFT.  We see from~\cite{Bonora} that it indeed exists. Natural object
  for this dynamics (as this paper will show) is a two-vertex. Again, we
  want to make Neumann coefficients of the two-vertices varying with the
  $t_k$ or (to put it differently) depending on the arbitrary conformal
  maps. What can be the meaning of such an object and where can it
  naturally appear?
  
  Let us contract a surface state with Witten's three-vertex. According to
  ``gluing theorem''~\cite{ss} the result will be some two-vertex.  This
  two-vertex can be considered as a \emph{multiplication operator}. For a
  given surface state $\ket \Sigma$ this operator $\hat \Sigma$ is defined in
  the following way
\begin{displaymath}
  \hat \Sigma: \ket X \to \ket{X*\Sigma}\quad \forall\: \ket X
\end{displaymath}
and can be realized via two-vertex $\ket{V_{12}(\Sigma)}$.  Again, this
two-vertex can be viewed as a functional on the whole family of surface
states. The resulting $\ket{V_{12}(t_k)}$ is Toda tau-function not for fixed,
but for dynamical $t_k$.

This object, which was just introduced following the intuition led by
integrable structure, has quite interesting interpretation by itself.  It
leads to a very intriguing possibility of describing solitonic objects of open
SFT as some boundary states.  Indeed, as we just said, some two-vertices can
be interpreted as boundary states~\footnote{Again, this observation, which can
  be made directly, originated from integrable structure -- from the fact that
  two-vertex and boundary state are expressed in terms of the same data,
  second derivatives of the same Toda tau-function}.  Thus multiplication
operators $\ket{V_{12}(\Sigma)}$ will in fact be boundary states for some
special surface states $\ket\Sigma$.  A very interesting problem is to find a
criteria the surface state should satisfy to give rise to a boundary state.
We do not try to address this question in its full mathematical generality,
but rather focus on various physically relevant examples.

It is a plausible assumption that some of these boundary states have
relations to D-branes. Thus, as a first example of such surface states in SFT,
we come to vacuum SFT (VSFT) (see, e.g.~\cite{vsft}) where D-branes as
solitonic solutions of equations of motion were conjectured~\cite{SenBr}.  In
the framework of VSFT these solutions are described via star-algebra projector
states -- they square to themselves under star-product~\cite{star-alg}.  Their
identification as branes is based on the fact that they yield correct ratios
of tensions of branes of different dimensions.  We take one of such
projectors, which evaded an interpretation so far -- the identity state -- and
find that it does give rise to orientifold boundary state. We present some
evidence that this should be also true in case of \emph{sliver state}. In
general we conjecture that the solutions identified in~\cite{SenBr} with
D-branes lead to Dirichlet and Neumann boundary states. If true, this would
not only allow to test the conjecture of~\cite{SenBr}, but also provide a new
candidate for description of closed strings in the open string field theory.

There is a variety of results which could be easily obtained once the
integrability structure of the problem is identified.  Because $t_k$'s
parameterize the whole family of surface states and surface states form a
sub-algebra under star product it should be possible to re-write star
multiplication analytically in terms of $t_k$. This would allow us, for
example, use the well-known integrable reductions to identify the
finite-dimensional sub-algebras of star-product.  Another possible development
can be made by giving sense to derivatives of Neumann coefficient w.r.t.\ 
$t_k$, i.e.\ the third derivatives of tau-function.  In~\cite{wwdvv} it was
shown that associativity (WDVV) equations~\cite{wdvv} are solved by
tau-function as a direct consequences of Hirota identities. The integrable
structure of SFT, identified in this paper, thus strongly suggests the
presence of such associativity algebra. The identification of the chiral rings
of associative operators in SFT is just one of the numerous possibilities
opened by integrability.

The paper has two logical parts.  We start by introducing the setup and
showing how conformally transformed vacuum of scalar CFT is related to the
holomorphic (anti-holomorphic) part of dToda tau-function in
Section~\ref{sec:cft-vac}.  (The corresponding material about (dispersionless)
Toda Lattice hierarchy is reviewed in Appendix~\ref{app:toda}).  This state
has an interpretation as a surface state of OSFT.  We show that Neumann
coefficients of such surface state satisfy Hirota identities.
In Section~\ref{sec:bcft} we consider the free scalar CFT with imposed Neumann
(Dirichlet) boundary conditions on an arbitrary analytic curve \CC. We show
that it is just a conformally transformed Neumann (Dirichlet) boundary state on
a circle.  We relate the boundary state on a curve \CC\ to the whole dToda
tau-function (not only to the holomorphic part of it).  
We review the basic objects of SFT, two- and three-vertices in
Section~\ref{sec:2vertex}.  The two-vertex gets identified with a Neumann
boundary state on a curve.  Having done that, we are able to relate the
two-vertex to the same Toda tau-function, it is done in
Section~\ref{sec:ident-with-toda}.  In Section~\ref{sec:3vertex} we comment on
the relation between three-vertex and dToda tau-function.  

Sections~\ref{sec:surf-bs}--\ref{sec:discus} comprise the second part of the
paper, addressing the issue which can be considered independently of the story
with integrability. Although the motivation comes from the problem of finding
dynamical interpretation for times $t_k$ discussed above.  In
Section~\ref{sec:surf-bs} we investigate the consequences of the
correspondence between two-vertex and a boundary state.  We conjecture a way
to build boundary states that correspond to solitonic solutions of certain
VSFT and explain how these boundary states would allow to test conjecture
of~\cite{SenBr}. We explore this correspondence first on the sub-algebra of
\emph{wedge states}. This is done in the Section~\ref{sec:wedge}.
Section~\ref{sec:ident} contains an explicit example of the correspondence, we
show in it that the identity state of the star-algebra can be identified with
the orbifold boundary state. This triggers some additional comments, presented
in the Section~\ref{sec:inv}.  We discuss all the open issues in the
Section~\ref{sec:discus}.

All technical details are given in the appendices. Appendix~\ref{app:toda}
contains a digest of various topics related to integrable hierarchies.  It is
intended to give a reader who is unfamiliar with the subject a quick overview
of KP and Toda Lattice hierarchies as well as their dispersionless limits;
introduces basic notions and approaches, and provide (non-exquisite) list of
references.  We give a proof of relation between Neumann boundary state on a
circle and a two-vertex of SFT in Appendix~\ref{app:2vert-circ}.  In
Appendix~\ref{app:2vert-conf} two-vertex of SFT defined by arbitrary conformal
transformations is related to the a boundary state on an analytic curve.
Finally, in Appendix~\ref{app:2vert-toda} we give a full list of
identification between the two-vertex and tau-function of Toda Lattice
hierarchy.

\section{Vacuum state in CFT and dToda tau-function} 
\label{sec:cft-vac}

Consider a scalar field defined on a whole complex plane.  It is a sum of
independent holomorphic and anti-holomorphic
components\footnote{
  \label{fn:1}%
  For field $\phi(w,\bw)$ to be single-valued in the complex plane one should
  impose the condition
   \begin{displaymath}
    \label{eq:1}
    b_0 = \bb0
  \end{displaymath}
  However it is convenient for us to keep both of these operators for now.
}
\begin{equation}
  \label{phi}
    \phi(w,\bar w) =\phi_0 + \phi(w) + {\bar \phi}(\bar w) = \phi_0+\hb0
    \log w +\bb0 \log {\bar w}-
    \mathop{\sum\nolimits'}\limits_{k=-\infty}^\infty\left(\frac{\hb k}{ k\, 
    w^k}+\frac{\hbt k}{ k\, {\bar w}^k} \right)
\end{equation}
The operators $b_k,\bb k$ 
are harmonics of two independent currents $J(w)=\de \phi(w,\bw)$,
${\bar J}(\bw) = \bar\de  \phi(w,\bw)$
\begin{eqnarray}
b_k = \oint_\infty \frac{dw}{2\pi i} w^{k} J(w)~~~~
\bb k = \oint_\infty \frac{d\bw}{2\pi i} \bw^{k} {\bar J}(\bw)
\end{eqnarray}
After quantization they obey the usual commutation relations
\begin{eqnarray}
[b_k,b_n] = [\bb k,\bb n] = k \delta_{k,-n}
\end{eqnarray}
One can construct a Fock space which is a tensor product of Fock
spaces for holomorphic and antiholomorphic sectors correspondingly, 
 in which  operators $b_k$ and $b_{-k}$ ($\bb{k}$ and $\bb{-k}$) act as
 annihilation and creation operators. 
The vacua in these Fock spaces are defined by
\begin{eqnarray}
  \label{vac}
  b_k \ket{p_b} = 0,\quad\forall\: k\geq 1 \\
  \bb k \ket{{\bar p}_b} = 0,\quad\forall\: k\geq 1 
  \label{eq:2}
\end{eqnarray}
where $p_b$ (${\bar p}_b$) are eigenvalues of the operators $b_0$ (\bb0) (see
the footnote~\ref{fn:1}).
Since $\phi(w,\bw)$ is a scalar field, currents $J(w)$
and ${\bar J}(\bw)$ are  primary fields with  conformal dimensions
$\Delta=(1,0)$, $\bar \Delta=(0,1)$ correspondingly.  Thus under conformal
transformation $(z,\bar z)\to (w(z),\bar w(\bar z))$ they change as
\begin{eqnarray}
\label{eq:3}
J(w) = \frac{dz}{dw} {\cal J}(z) \\
{\bJ}(w) = \frac{d{\bar z}}{d{\bar w}} {\bar {\CJ}}({\bar z}) 
\end{eqnarray} 
These transformations can also be written in terms of operators $U_w$
representing elements of Virasoro group
\begin{equation}
  \label{eq:4}
  U_w = \exp(\sum_{n} v_n L_n)
\end{equation}
where $v_n$'s are harmonics of $v(z) = \sum v_n z^{n+1}$ 
and field $v(z)$ is defined by equation
\begin{equation}
  \label{eq:5}
  e^{v(z)\p_z} z = w(z)
\end{equation}
For the purpose of this paper let us constrain ourselves to the univalent
transformations, i.e. those that map $z=\infty$ into $w=\infty$ in one-to-one
manner. Then $w(z)$ has the form
\begin{eqnarray}
\label{eq:6}
w(z) = \frac{z}{r} + \sum_{k\geq 0} \frac{p_k}{z^k}
\end{eqnarray}
Here we note that $r$ as well as other parameters are complex.
The $b_k$'s can be expressed through the conformally
transformed current ${\cal J}(z)$ as 
\begin{eqnarray}
\label{eq:7}
 b_k = U_w^{-1} a_k U_w = \oint \frac{dz}{2\pi
  i} \Big(w(z)\Big)^{k} {\CJ}(z) 
\end{eqnarray} 
and similarly for $\bb k$'s. Here  $a_k$'s are  harmonics 
of the current $\CJ(z)$.
For univalent transformations (\ref{eq:6}) this means for $b_k$'s
\begin{equation}
  \label{eq:8}
  \begin{array}[c]{lcll}
    b_n& =& \displaystyle\sum_{k=0}^{n} C_{n,k} a_k + \sum_{k=1}^{\infty}
    C_{n,-k} 
    a_{-k}&\forall n>0\\
    b_{-n} &=& \displaystyle\sum_{k=n}^{\infty} C_{-n,-k} a_{-k}&\forall n>0 
  \end{array}
\end{equation}
\begin{equation}
  \label{eq:9}
  \begin{array}[c]{lcll}
    C_{n,k}&=& \displaystyle\oint \frac{dz}{2\pi i} z^{-k-1} \Big(w(z)
    \Big)^n&\forall k,n 
  \end{array}
\end{equation}
By expressing $b_k$'s via $a_k$'s, we establish the map from the Fock
space of the former operators to that of the latter. 
The natural question to ask is how the
image of the vacuum~(\ref{vac}--\ref{eq:2}) behaves under this map. 
Let $\ket{p_a}$ and $\ket{{\bar p}_a}$ be vacua
in the Fock spaces of operators $a_k$ ($\bar a_k$).
They are defined via equations similar to~(\ref{vac}--\ref{eq:2})
but with respect to operators $a_k$. Since 
$b_k(a) = U_w^{-1} a_k U_w$ the state 
\begin{equation}
\label{eq:10}
\ket{w}\otimes\ket{{\bar w}} = U_w^{-1} \ket{p_a}
\otimes {\bar U}_{\bar w}^{-1} \ket{{\bar p}_a}
\end{equation}
in the Fock space of operators $a_k$ ($\ba k$) satisfies
\begin{eqnarray}
  \label{eq:11}
  b_k(a) \ket{w} =\bb{k}(\ba{}) \ket{{\bar w}}  = 0 ~~~~\forall k\geq 1
\end{eqnarray}
We will call the state $\ket w\otimes\ket{\bar w}$ \emph{conformally transformed vacuum}.
Since the relation between
$b_k$'s and $a_k$'s is linear we can look for the solution of eq.~(\ref{eq:11})
in the form
\begin{eqnarray}
\label{eq:12}
\ket{w}\otimes\ket{{\bar w}} = \exp{\left( \frac{1}{2}
\sum_{k,m=0}^{\infty} a_{-k}a_{-m} N_{km} \right)} \ket{p_a}
\otimes
\exp{\left( \frac{1}{2}
\sum_{k,m=0}^{\infty} \ba{-k}\ba{-m}\bN_{km} \right)}
\ket{{\bar p}_a}
\end{eqnarray}
The coefficients $N_{km}$ are subject to constraints, following from
equations 
\begin{eqnarray}
  \label{eq:13}
  b_n \ket{w} 
  =  \left[ \sum_{k=0}^{n} C_{n,k} a_k + \sum_{k=0}^{\infty} C_{n,-k} a_{-k}
    \right] 
  \exp{\left( \frac{1}{2}
    \sum_{i,j=1}^{\infty} a_{-i}a_{-j} N_{ij} \right)} \ket{p_a} = 0
\end{eqnarray}
Thus
\begin{eqnarray}
\sum_{k=1}^{n} k C_{n,k} N_{km} + C_{n,-m} = 0~,~~~m\geq 0,n\geq 1
\label{vacuum.trans} 
\end{eqnarray}
One can show (see Appendix A in~\cite{Bo-Ru}) that these equations are solved by
\begin{eqnarray}
\label{Nkm}
N_{km} &=& \frac{1}{km}\oint \frac{dz}{2\pi i} \oint \frac{d\zeta}{2\pi i}
z^k \zeta^m \de_z \de_\zeta\log\Big( w(z) - w(\zeta) \Big) \\
N_{0m} &=& \frac{1}{m}\oint \frac{dz}{2\pi i} z^m \de_z \log\Big( w(z)\Big) \nonumber
\end{eqnarray}
Coefficients $\bN_{km}$ depend on ${\bar w}$ in the same way. 
The above result can be checked by direct calculations. 

Solution~(\ref{Nkm}) requires some comments. For any function $w(z)$ regardless
of its interpretation as a conformal map~(\ref{eq:6}) coefficients
$N_{kn},\;k,n\ge0$ defined via~(\ref{Nkm}) \emph{are not
  independent}, but rather obey an infinite set of relations of which the
first few are
\begin{equation}
\renewcommand{\arraystretch}{1.2}
  \label{eq:14}
  \begin{array}[c]{rcl}
  N_{22} &=&  N_{13} - \dfrac 12 N_{11}^2 \\
  N_{23} &=&  N_{14} - N_{11}N_{12}\\
  N_{33} &=& \dfrac 13 N_{11}^3 -  N_{11} N_{13} - N_{12}^2 +  N_{15} \\
  &\cdots&
\end{array}
\end{equation}
It is, of course, possible (although not obvious) to see these relations
directly from eq.~(\ref{Nkm}). To appreciate how non-trivial they
are one may want to take a look at the example of Neumann matrix in,
say,~\cite{zw}, eq.~(2.14).

Note that the answer~(\ref{Nkm}) was obtained in~\cite{peskin,peskinII} from
a different point of 
view. It was shown there that coefficients $N_{km}$ are harmonics
of conformally transformed scalar field propagator
\begin{eqnarray}
\label{eq:15}
\langle J(w(z)) J(w(\zeta))\strut\rangle = \de_z \de_\zeta \log\Big( w(z) - w(\zeta)\Big)
\end{eqnarray}
However, it was absolutely not obvious that one should search for the
relations of the type~(\ref{eq:14}) in the setup of~\cite{peskin}. And indeed,
they were not found.

To see the nature of these relations one should recognize an underlying
structure of conformally transformed vacuum. It was shown in~\cite{Bo-Ru}
that one can
introduce the generating function $F(\t0,\t k,\tb k)$ such that
\begin{equation}
\label{eq:16}
N_{km} = \frac{1}{km} \pfrac{^2 F}{t_k \p t_m},\quad N_{k0} = \frac{1}{k}
\pfrac{^2 F}{t_k \p t_0},\quad N_{\bar k\bar m} = \frac{1}{km} \pfrac{^2
  F}{\bt k \p \bt m},\quad N_{\bar k0} = \frac{1}{k}
\pfrac{^2 F}{\bt k \p t_0}
\end{equation}
By substituting~(\ref{eq:16})
into~(\ref{vacuum.trans}) the latter becomes equivalent to the set of equations
\begin{equation}
\label{eq:17}
  (z-\zeta)e^{D(z)D(\zeta)F} = ze^{-\partial_{\t0}D(z)F} - \zeta e^{-\partial_{\t0}D(\zeta)F}\
\end{equation}
(for definition of operator $D(z)$ as well as all other notations and short
review of Toda Lattice Hierarchy see Appendix~\ref{app:toda}, particularly
section~\ref{sec:disp-limit-toda}).  
One recognizes in~(\ref{eq:17}) \emph{Hirota equations of dispersionless Toda
  Lattice Hierarchy}. This means, that function $F$ defined in~(\ref{eq:16})
is the (logarithm of) tau-function of this hierarchy. We may define function
$w(z)$ as generating function for the second derivatives $\p_{\t0}\p_{t_k}F$
\begin{equation}
  \label{eq:18}
  D(z)\p_{\t0} F = -\log\frac{w(z)}{z/r}
\end{equation}
Eqs.~(\ref{eq:16}) and~(\ref{eq:18}) are equivalent to the equations for
$N_{0m}$ in~(\ref{Nkm}), if functions $w(z)$ in both of them are the same.
Then we can rewrite~(\ref{eq:17}) in the form
\begin{equation}
  D(z)D(\zeta)F - \frac{1}{2} \de^2_{t_0}F = 
  \log \frac{w(z) - w(\zeta)}{ z - \zeta } \label{eq:19}
\end{equation}
Expanding l.h.s. of eq.~(\ref{eq:19}) in $z^{-1}$, $\zeta^{-1}$ we obtain 
eqs.~(\ref{Nkm}) for $N_{km}$. 

By expressing from~(\ref{eq:17}) derivatives $\p_{\t0}\p_{\t k}F$ via
$\p_{\t1}\p_{\t k}F$ and substituting it back, Hirota eqs.~(\ref{eq:17}) can
also be rewritten in the \emph{pure holomorphic form}, which is also called
\emph{Hirota equations of dispersionless KP hierarchy}~\cite{kodama}
\begin{equation}
  \label{eq:20}
  \exp(D(z)D(\zeta)F) = 1 - \frac{D(z)\p_{\t1}F-D(\zeta)\p_{\t1}F}{z-\zeta}
\end{equation}
Expanding it in $z^{-1}$, $\zeta^{-1}$ and
using identifications~(\ref{eq:16}) we get 
equations on coefficients $N_{km}$. They are nothing else but 
equations~(\ref{eq:14}).

\section{Boundary states in CFT  and dToda tau-function}
\label{sec:bcft}

In the previous Section we showed that interpretation for the purely
(anti)holomorphic second derivatives of the dispersionless tau-function
$F(\t0,\t k \tb k)$ of
dToda  hierarchy can be found in terms of the Neumann coefficients of
the conformally transformed vacuum states. The natural question would be
whether there is a similar interpretation for the mixed
(holomorphic-antiholomorphic) derivatives of this function  (see
Appendix~\ref{sec:disp-limit-toda} for details). To find such interpretation we
would have to find a state in the Fock space of scalar field, which mixes
holomorphic and anti-holomorphic Fock spaces of its components.

Consider again the scalar field in the plane $w$ (c.f. eq.~(\ref{phi}))
\begin{equation}
  \label{eq:21}
    \phi(w,\bar w) = \phi_0+\hb0
    \log w +\bb0 \log {\bar w}-
    \mathop{\sum\nolimits'}\limits_{k=-\infty}^\infty\left(\frac{\hb k}{ k\, 
    w^k}+\frac{\hbt k}{ k\, {\bar w}^k} \right)
\end{equation}
and impose the Neumann
\begin{equation}
  \label{eq:69-n}
 \partial_{\mathrm{n}} \phi(w,\bar w)\Bigr|_{|w|=1} = 0
\end{equation}
or Dirichlet 
\begin{equation}
  \label{eq:22}
  \phi(w,\bar w)\Bigr|_{|w|=1} = 0
\end{equation}
boundary conditions on the unit circle.  Presence of the boundary makes
holomorphic and anti-holomorphic modes dependent. As it is well known, there
are two ways to realize it in quantum theory.  One can either solve boundary conditions at the
classical level: $\bb{} = f(b)$ (``open string picture'') and therefore
express explicitly $\bar b$ in terms of $b$ or quantize the theory first
and then impose boundary conditions as constraints on quantum states
(``closed string picture''). The latter leads to boundary state
construction~\cite{BS}. Boundary state $\bket{B}$ is defined
by
\begin{eqnarray}
\label{eq:bs}
\bra{0}V(b,\bb{}){\biggr|_{\bar b =
  f(b)}}\ket{0}_{open}=\bra{0}V(b,\bb{})\bket{B}_{closed} 
\end{eqnarray}
The boundary conditions~(\ref{eq:22}) or~(\ref{eq:69-n}) imply
constraints $b_n \pm {\bar b}_{-n} = 0$. In ``closed string picture'' 
they should be imposed on states.
In  the present simple case of the free scalar theory they define the boundary
state uniquely via
\begin{equation}
  \label{eq:23}
   (\hb{n} \pm  \bb{-n}) \bket{B}_{circle}=0,\;\forall\:n
\end{equation}
where upper sign ($+$) corresponds to Neumann boundary conditions and lower
sign ($-$) to Dirichlet conditions.  
In what follows we are going to work mostly with the case
of Neumann boundary state. We can solve~(\ref{eq:23}) explicitly
\begin{eqnarray}
\label{eq:24}
\bket{N_b}_{circle} = \int dpd\bar p \,\delta(p+\bar p)
\exp\left\{- \sum_{k>0} \frac{b_{-k}{\bar
      b}_{-k}}{k}\right\} \ket{p}\otimes\ket{\bar p}
\end{eqnarray}
We would like to emphasize that for the case of Neumann boundary state
single-valuedness condition $p=\bar p$ (see the footnote~\ref{fn:1},
p.~\pageref{fn:1}) together 
with the delta-function in the above equation 
selects $p=\bar p =0$ as the only choice. However, for
the purpose of this paper, we will not impose single-valuedness here.

Now we would like to find Neumann boundary state $\bket{N}_\CC$ for the
boundary conditions imposed on an
arbitrary analytic curve $\CC$. This can be done applying conformal
transformations in the way analogous to Section~\ref{sec:cft-vac}.  In
accordance to Riemann mapping theorem there exists a conformal transformation
$w(z)$ that maps any analytic curve ${\cal C}$ to a unit circle and exterior
of the curve into exterior of the unit circle.  If a scalar field $\phi(w,\bar
w)$ satisfies Neumann (Dirichlet) boundary conditions on the circle, then
conformally transformed field $\phi(w(z),\bar w(\bz))$ satisfies the same
condition on the curve \CC. Since $b_n=U^{-1}_w a_n U_w$ (and analogous for
antiholomorphic sector)\footnote{See eqs.~(\ref{eq:4}--\ref{eq:7}) for
  details} the boundary conditions (\ref{eq:23}) imply the following for
$\bket{N}_\CC$
\begin{equation}
\label{eq:25}
\Biggl( U^{-1}_w a_n U_w + {\bar U}^{-1}_{\bar w} {\bar a}_{-n} {\bar U}_{\bar
  w} \Biggr) \bket{N}_\CC = 0
\end{equation}
Here $U_w$ and ${\bar U}_{\bar w}$ act on holomorphic/antiholomorphic
sectors of $\bket{N}_\CC$. It is clear that
\begin{equation}
\label{eq:4.1}
\bket{N}_\CC =  U^{-1}_w {\bar U}^{-1}_{\bar w} \bket{N_a}_{circle}
\end{equation}
is a solution of (\ref{eq:25}) (compare to conformally transformed vacuum in
eq.~(\ref{eq:10})).
Here $\bket{N_a}_{circle}$ is defined by equations similar to~(\ref{eq:23}),
but in the Fock space of operators $a_k$, $\ba k$.

We proceed to finding this solution. After plugging in the eqs.~(\ref{eq:23})
$b_k(a)$ 
given by eqs.~(\ref{eq:8}), one sees that the state $\bket{N}_\CC$ obeys
\begin{equation}
  \label{eq:26}
 \left[\,\sum_{k=0}^n C_{n,k}\ha k  + \sum_{k=1}^\infty
   C_{n,-k} \ha{-k} +  \sum_{k=n}^\infty {\bar C}_{-n,-k} \at{-k}\right]
   \bket{N}_\CC= 0 
\end{equation}
where $C_{k,n}$ are given by eq.~(\ref{eq:9}). As before we 
try to solve~(\ref{eq:26}) as
\begin{eqnarray}
\label{eq:27}
\begin{array}[c]{rl}
  \bket{N}_\CC &=  \displaystyle\int dp\,d\bar p\,\delta(p + \bar p)\times\\
&\exp\left\{\displaystyle
\sum_{k,n\ge0} B_{k{\bar n}} a_{-k}{\bar a}_{-n}
+ \frac12\sum_{k,n\ge0} B_{kn} a_{-k} a_{-n} 
+ \frac12\sum_{k,n\ge0} B_{{\bar k}{\bar{\vphantom{k} n}}} {\bar a}_{-k}
{\bar a}_{-n} \right\}\ket{p}\otimes\ket{\bar p}
\end{array}
\end{eqnarray}
Substituting~(\ref{eq:27}) into (\ref{eq:26}) we find that the coefficients
$B_{kn}$ obey the same equations~(\ref{vacuum.trans}) as $N_{kn}$ for
conformally transformed vacuum state (and analogs thereof for $B_{{\bar
    k}{\bar{\vphantom{k} n}}}$ with coefficients being conjugated and $a_k$
interchanged with $\ba k$).  Thus the solutions for them are the Neumann
coefficients~(\ref{Nkm}).  As for expressions $B_{k{\bar n}}$ with mixed
indices, they can be extracted from a condition similar to~(\ref{vacuum.trans})
\begin{equation}
  \label{eq:28}
  \sum_{k=1}^n k C_{n,k}
    B_{k{\bar m}} + {\bar
    C}_{-n,-m}=0
\end{equation}
Due to the fact that expressions for $B_{kn}$ and $B_{{\bar
    k}{\bar{\vphantom{k} n}}}$ are the same as in Section~\ref{sec:cft-vac} (and
can therefore be identified with second derivatives of tau-function $F$), we may try
the following identification for $ B_{k{\bar n}}$
\begin{equation}
  \label{eq:29}
   B_{k{\bar n}} = \frac{1}{kn} \pfrac {^2 F}{\t k\p {\tb n}}
\end{equation}
Again, one can show (see~\cite{Bo-Ru}, Appendix~B) that equation~(\ref{eq:28})
together with identification (\ref{eq:29}) is equivalent to Hirota
equations~(\ref{eq:122}). Thus the identification gives us
\begin{eqnarray}
  \label{eq:30}
  B_{k{\bar n}} = - \frac{1}{kn} 
  \oint \frac{dz}{2\pi i} \oint \frac{d\bz}{2\pi i} 
  z^k \bz^n \de_z \de_{\bar z} \log
\left( 1 - \frac{1}{w(z) {\bar w}(\bz) } \right)
\end{eqnarray}
We present the final result for Neumann (upper sign) and Dirichlet (lower sign)
boundary state written in terms of derivatives of tau-function
\begin{equation}
  \label{eq:31}
  \begin{array}[c]{rcl}
    \bket{B}_\CC &=& \displaystyle\int dp\exp\left\{\dfrac{p^2}2
      \dfrac{\p^2 F}{\p\t0^2}+\frac12\sum_{k,n=1}^\infty\left(a_{-k}
        a_{-n}\dfrac{\p^2 F}{\p \t k \p \t n}  + \ba {-k} \ba{- n}\frac{\p^2 F}{\p
          \tb k \p \tb n}\right)+\right.\\[16pt]
    &+&  \left . p\sum\limits_{k=1}^\infty \left(a_{-k} \dfrac{\p^2
          F}{\p\t0\p\t k} \mp 
        \ba{- k} \frac{\p^2 F}{\p\t0\p\tb k}\right) \mp
      \sum\limits_{k,n=1}^\infty  a_{-k} \ba {-n}\dfrac{\p^2 F}{\p \t k \p
        \tb n}\right\}\ket{p}\otimes\ket{\mp p}
\end{array}
\end{equation}
here we have fixed the normalization factor to be 
$\frac{p^2}2 \frac{\p^2 F}{\p\t0^2}$, 
so that the state $\bket{D}$ would be the generating
function for \emph{all} second derivatives of dispersionless 2D Toda
tau-function.

In the case of Dirichlet boundary state, eq.~(\ref{eq:31}) has an interesting
interpretation. If we realize operators $a_{-k}$ and $\ba{-k}$ as operators of
multiplications on $s_k$ and $\bar s_k$, then the 
state $ \bket{D}_\CC$ as a 
function of $s_k$, $\bar s_k$ will itself be a tau-function of dispersionless
Toda hierarchy  (see~\cite{Bo-Ru} for details). Neumann boundary state,
being quite similar in construction, lacks this interpretation to the best of
our knowledge. Namely, the difference which prevents such interpretation is
the following. First of all, relative minus sign in front of mixed derivative
(compared to (anti)holomorphic sectors) appears in the case of Neumann
boundary conditions. Secondly, in this case $p=0$ if we require that scalar
field~(\ref{phi}) is single valued. Under such condition terms with
derivatives with respect to $\t0$ would disappear. 

Another comment which should be made about~(\ref{eq:31}) is that we never used
the fact that $\t k$ and $\bt k$ are complex conjugated. This also implies
that in general $w(z)$ and $\bar w(\zeta)$ in above formulae can be independent
rather than complex conjugated. Eq.~(\ref{eq:31}) will work formally for this
general case as well, although the physical interpretation of $\bket{B}_\CC$
is less clear. We will come back to this point at the end of
Section~\ref{sec:2vertex}.

\section{Two-vertex of SFT as a boundary state}
\label{sec:2vertex}

In Section~\ref{sec:cft-vac} we have already mentioned that the
ansatz~(\ref{Nkm}) (without identification with Toda) was obtained
in~\cite{peskin,peskinII} in the context of open SFT. At first glance there
can not be any connection of $B_{k\bar n}$ (defined
via~(\ref{eq:27})--(\ref{eq:29})) with SFT, because these coefficients
appeared in the expression for the boundary state, whereas in open string
field theory one works in the ``open string picture'' and there is no place
for boundary state as such.  Nevertheless, recently it was
noticed~\cite{Bonora} that mixed derivatives of dispersionless Toda
tau-function (at particular values of its arguments) do appear in the description
of SFT three-vertex. In this Section we will show that it is not a
coincidence. To wit, there is a close relation between boundary states and SFT
vertices. Using this relation we will derive the identification between
dispersionless Toda tau-function and Neumann coefficients in SFT in a
systematic way.

Let us remind the basic facts about the \emph{Cubic String Field Theory}.
Action of this String Field Theory has the following schematic
form~\cite{wittenSFT}
\begin{equation}
  \label{eq:32}
  S_{SFT} = \frac12\int \Phi*Q_{BRST}\Phi +\frac13\int \Phi*\Phi*\Phi
\end{equation}
We will not discuss the kinetic term here.  The $*$-product in~(\ref{eq:32})
is the main defining object of SFT. It encodes all the interactions by
specifying how two of the incoming strings are glued into the resulting one.
All strings can be off-shell.  Thus star-product is a map from tensor
product of two string Hilbert spaces into that of a third one
\begin{eqnarray}
\label{eq:33}
*:\:\CH_1 \otimes \CH_2\to \CH_3
\end{eqnarray}
One way to specify $*$-product would be to introduce a
\emph{three-vertex} $\ket{V_{123}}\in\CH_1\otimes\CH_2\otimes\CH_3$, such that
\begin{equation}
\label{eq:34}
\ket{A*B}\equiv  \bra{A}\otimes\bra{B}\ket{V_{123}}
\end{equation}
where by $\bra A$ we mean a \emph{BPZ conjugation} of $\ket A$.
Then interaction term in~(\ref{eq:32}) can be written as
\begin{equation}
  \label{eq:35}
  \int \Phi_A*\Phi_B*\Phi_C= \sprod{C}{A*B}
\end{equation}
In order to define this three-vertex three in general independent conformal
transformations can be used
\begin{equation}
  \label{eq:36}
  \bra{0} h_1(V_A) h_2(V_B) h_3(V_C) \ket{0} =
  \bra{A}\otimes\bra{B}\otimes\bra{C}\ket{V_{123}}
\end{equation}
where $V_A,V_B,V_C$ are vertex operators creating 
states $\ket{A}$, $\ket{B}$, and $\ket{C}$ correspondingly.  It should be
noted that all three vertex operators in the l.h.s. of~(\ref{eq:36}) act
\emph{on the same} Hilbert space.  Maps $h_I(z),I=1,2,3$ are fixed. They
contain in themselves all the information about $\ket{V_{123}}$. Different
choices of $h_I(z)$ will lead to different off-shell interactions in 
String Field Theory.  For example, the choice that leads to Witten's three-vertex is
\begin{eqnarray}
\label{eq:37}
h_1(z) = T^2\circ h(z),\quad h_2(z) = T\circ h(z),\quad h_3(z) = h(z)
\end{eqnarray}
where $h(z)$ is a conformal map that carries a unit circle into a $120^\circ$
wedge thereof and $T$ is a $120^\circ$ rotation
\begin{eqnarray}
h(z) = \LB\frac{z+i}{z-i}\RB^{2/3},\qquad T(z) = e^{2\pi i/3} z
\end{eqnarray}
The three-vertex $\ket{V_{123}}$ is defined as\footnote{
  Depending on $h_I(z)$ it is more natural to work with \emph{bra} or
  \emph{ket} vertices. Throughout this paper we will be working with the ket
  ones, which are BPZ conjugated as compared to, say,~\cite{peskin}. It means
  that our $h_I(z)$ and $\tilde h_I(z)$ in~\cite{peskin} are related as:
  $\tilde h_I(z) = h_I(-\frac1z)$.\label{fn:2}
}%
\begin{equation}
 \label{3vert}
  \ket{V_{123}} =\int \prod_{I=1}^3 dp_I\, (2\pi)\, \delta\LB\sum_{I=1}^3p_I\RB
  \exp{\LB \frac12 \sum_{I,J=1}^3 \sum_{m,n=0}^{\infty} a_{-n}^I a_{-m}^J
    N^{IJ}_{mn}\RB} 
  \ket{p_1} \otimes  \ket{p_2} \otimes  \ket{p_3} 
\end{equation}
Here each operator $a^I$ acts on its own vacuum $\ket{p_I}$, momenta $p_I$
are eigen values of zero-modes $a^I_0$. The values of \emph{Neumann
  coefficients} $N^{IJ}_{nm}$ depend only on conformal maps $h_I$
\begin{eqnarray}
\label{eq:38}
N_{km}^{IJ} &=& \frac{1}{km}\oint \frac{dz}{2\pi i} \oint \frac{d\zeta}{2\pi i}
z^k \zeta^m \de_z \de_\zeta\log\Big( h_I(z) - h_J(\zeta) \Big)\\
\label{eq:39}
N_{0m}^{IJ} &=& - \frac{1}{m}\oint \frac{dz}{2\pi i} z^m \de_z \log\Big(
h_I(\infty)-h_J(z)\Big)\\ 
\label{eq:40}
  N_{00}^{IJ}&=&\left\{
  \begin{array}[c]{ll}
    \log ([h'_I](\infty))& I=J\\
    \log (h_I(\infty)-h_J(\infty))& I\neq J
  \end{array}\right.
\end{eqnarray}
where 
\begin{equation}
  [h'](z) \equiv - \frac{d h(z)}{d 1/z}\label{eq:41}
\end{equation}
In the above we assumed that we are going to work with the expansion of
functions $h_I(z)$ around infinity\footnote{
  Eqs.~(\ref{eq:38})--(\ref{eq:40}) may seem to be ill-defined if $h_I(z)$'s have
  poles at infinity. In Appendix~\ref{app:2vert-circ} we show how to deal
  with such formulae in these cases.
}.
Note that $N^{II}_{kn}$ for $k,n>0$ are given by analogs of eq.~(\ref{Nkm}).

Along with the 
three-vertex~(\ref{eq:34}) one can in principle define $n$-vertex
$\ket{V_{1\dots n}}$ for arbitrary $n$ as
\begin{eqnarray}
  \label{eq:42}
  \bra{0}h_1(V_1) \dots h_n(V_n)\ket{0} =
  \bra{V_1}
  \otimes\dots\otimes \bra{V_n}\;\ket{V_{1\dots n}}
\end{eqnarray}
The simplest case is a one-vertex 
(also known as a \emph{surface state} $\ket{h_1}$), defined by
\begin{equation}
  \label{eq:43}
  \ket{h_1} = \exp{ \LB
    \frac12  \sum_{n,m=1}^{\infty} a_{-n} N_{nm}^{11}
    a_{-m} \RB} \ket{0}
\end{equation}
where $N_{kn}^{11}$ as functions of some conformal transformation $h_1(z)$ are
given by eq.~(\ref{eq:38}) for $I=J=1$. It is easy to see that this object
coincides with conformally transformed vacuum state $\ket w$~(\ref{eq:12}) of
Section~\ref{sec:cft-vac} for $p=0$ (thus the notation) and its Neumann
coefficients $N_{kn}$ are expressed via second derivatives of dKP
tau-function. Like in the case of Dirichlet boundary state (see discussion 
after eq.~(\ref{eq:31})), surface state itself is a tau-function of
dKP if one realized operators $a_{_k}$ as multiplication on
variables $s_k$. For discussion see~\cite{Bo-Ru}.

The next example is a two-vertex $\ket{V_{12}}$, defined by
\begin{eqnarray}
  \bra{0} h_1(V_1) h_2 (V_2) \ket{0} =\bra{V_1} \otimes \bra{V_2}\;\ket{V_{12}} 
 \label{2vert}
\end{eqnarray}
It can be written again in terms of Neumann
coefficients~(\ref{eq:38}--\ref{eq:40}) 
\begin{eqnarray}
  \ket{V_{12}} = \int dp_1\, d p_2\,2\pi\delta(p_1+p_2)\exp{ \LB
    \frac12\sum_{I,J=1}^2   \sum_{n,m=0}^{\infty} a^I_{-n} N^{IJ}_{nm}
    a^J_{-m} \RB} \ket{p_1}\otimes\ket{p_2} 
  \label{2vert.N}
\end{eqnarray}
We will also use a notation
\begin{equation}
  \label{eq:44}
  \ket{V_{12}}\equiv \ket{h_1,h_2}
\end{equation}
specifying explicitly on which conformal transformations the two-vertex
$\ket{V_{12}}$ depends.

By its very definition~(\ref{2vert.N}) the two-vertex $\ket{V_{12}}$ is
similar to the boundary state\footnote{
  If one identifies $a^1_k$ and $a^2_k$ in~(\ref{2vert.N}) with $a_k$ and $\ba
  k$ of Section~\ref{sec:bcft}. Note, however, that in case of two-vertex
  there is no additional condition requiring $p_1 = p_2$.\label{fn:3}%
}~(\ref{eq:bs}).  We would like to explore this similarity in details and find out {what}
boundary theory the two-vertex~(\ref{2vert.N}) corresponds to.

It is not hard to find explicit conformal maps that define the two-vertex state
which is Neumann boundary state on a unit circle.  The required conformal
maps are 
\begin{equation}
  \label{eq:45}
  h_1(z) = z, \quad h_2(z) = \frac1z
\end{equation}
Indeed, the Neumann coefficients~(\ref{eq:38}) for these maps are\footnote{
  Neumann coefficients $N^{IJ}_{mn}$ are invariant under \gl2
  transformations.  Thus we could consider
  \begin{displaymath}
    h_1(z) = \frac{a z + b}{c z + d},\quad    h_2(z) = \frac{(1/z) a +
      b}{(1/z) c + d}
  \end{displaymath}
  with $ad -bc \neq 0$.
   \label{fn:4}%
} 
(see Appendix~\ref{app:2vert-circ} for details)
\begin{eqnarray}
  \label{eq:47}
  &&N^{12}_{nm}=  N^{21}_{nm} = -\frac{1}{n} \delta_{n,m} \\
  \label{eq:48}
  &&N^{11}_{nm}= N^{22}_{nm}  = -\frac{1}{n} \delta_{n,-m}
\end{eqnarray}
Plugging~(\ref{eq:47}--\ref{eq:48}) into  definition of a vertex
state~(\ref{2vert.N}) we get
\begin{eqnarray}
\label{eq:49}
\ket{V_{12}} = 
 \int dp_1 \,dp_2\,2\pi\delta(p_1+p_2)\exp{ \left( -\sum_{n>0} \frac{a^1_{-n}
       a^2_{-n}}{n} \right)}\ket{p_1}\otimes \ket{ p_2}
\end{eqnarray}
Eq.~(\ref{eq:49}) coincides with the boundary 
state on a unit circle~(\ref{eq:24})
(see the footnote~\ref{fn:3} \vpageref{fn:3}). As it has already been mentioned there this would
become the usual Neumann boundary state if we imposed single-valuedness
condition. Note, that for two-vertex state there is no a priori reason to do
it.

We can finally establish the relation between a two-vertex defined by
arbitrary maps and Neumann boundary state on an arbitrary curve
$\bket{N}_\CC$.  Let function $w(z)$ map exterior of this curve to the
exterior of a unit circle.  To differentiate between this conformal map and
conformal maps defining two-vertex we still denote the latter by
$h(z)$. In Appendix \ref{app:2vert-conf} we show that
\begin{equation}
  \label{eq:50}
  \ket{h_1,h_2}=\bket{N}_\CC
\end{equation}
where
\begin{equation}
  \label{eq:51}
    h_1(z) =  w(z),\quad  h_2(z) = \frac1{\bar w(z)}
\end{equation}
or some \gl2 transformation thereof\hphantom{\,}\footnote{
  Because of \gl2 freedom in definition~(\ref{eq:45}) mentioned before
  (the footnote~\ref{fn:4}~\vpageref{fn:4}) the same ambiguity is present
  in~(\ref{eq:51}).  However, requiring specific analytic properties of $w(z)$
  and $h_I(z)$, this freedom can be completely fixed. For details see
  Appendix~\ref{app:2vert-conf}.\label{fn:5}}.
In case of the circle $w(z)=\bar w(z) = z$ and~(\ref{eq:51}) becomes~(\ref{eq:45})
as it should.  Note, that given any analytic curve one can construct a
two-vertex $\ket{w,\bar w^{-1}}$ out of it.  Inverse is not true. Given
arbitrary $h_1$ and $h_2$ one can not always find an appropriate analytic curve
(obtained via~(\ref{eq:51}) $w(z)$ and $\bar w(z)$ will not be complex
conjugated). Thus, two-vertex may be thought of as generalization of the
boundary state.

To understand the meaning of eq.~(\ref{eq:51}) 
let us come back to eq.~(\ref{eq:50}).
Recall, that by its definition~(\ref{2vert}) two-vertex
$\ket{h_1,h_2}$ belongs to the tensor product of Hilbert spaces of two
\emph{open strings} and one can think of each open string as living in the
exterior of a unit circle in its own complex plane\footnote{
  We are working with the \emph{ket}-states, thus natural picture for the open
  string world sheets are exteriors of various curves, e.g. unit  circles.}. 
At the same time, boundary state in the r.h.s. of~(\ref{eq:50}) belongs to the
Hilbert space of closed strings. In the simplest case~(\ref{eq:45}) world
sheets of the two open strings are represented by complex planes with
removed unit disks. Then transformation $h_1(z)=z$ maps (identically) this
world sheet onto the exterior of the unit circle, while $h_2(\bz)=1/\bz$ maps
world sheet of the second string onto the interior of the unit circle.  The
reason why we write here $h_2(\bz)$ instead of $h_2(z)$ is the following.
Eq.~(\ref{eq:45}) should be actually written as $h_1(z_1)$ and $h_2(z_2)$
 to stress that functions $h_1$
and $h_2$ are defined on different spaces. When we glue two world sheets of the
open strings into the one world sheet of the closed string we want maps $h_1$
and $h_2$ to match continuously across the boundary (unit circle).  This
means that we 
should identify $z_1$ with $z$ and $z_2$ with $\bz$ (then $h_1(z) = h_2(\bz)$
when $z \bz = 1$)\footnote{Compare this with the comment in the
footnote~\ref{fn:3} \vpageref{fn:3}.}. 
 Under this
identification two-vertex~(\ref{eq:49}) is related to Neumann boundary state
on the unit circle~(\ref{eq:24}). This means that a correlator in the r.h.s.
of~(\ref{2vert}) is computed in the closed string picture with Neumann
boundary state on the unit circle.  

More general situation~(\ref{eq:51}) is obtained if one considers open strings
with Neumann boundary conditions on an arbitrary analytic curve \CC. Now the
world sheet of the open string is a complex plane with the interior of the
curve $\CC$ removed. In order to combine two such open strings into a closed
string as required by eq.~(\ref{eq:50}) we should map the exterior of the
curve $\CC$ (world sheet of the first string) into the exterior of the unit
circle on the world sheet of the closed string and the world sheet of the
second string into the interior of the same unit circle.  The transformations
$h_1(z)$ and $h_2(\bz)$ of eq.~(\ref{eq:51}) do precisely that.

\subsection{The identification of two-vertex with dToda tau-function}
\label{sec:ident-with-toda}

In Sections~\ref{sec:bcft} and \ref{sec:2vertex} we were able to relate Neumann
boundary state on an arbitrary analytic curve to two different
objects - dispersionless Toda tau-function and two-vertex of SFT. 
That shows that there is a close connection between the two in their own right.
Namely, one can express a two-vertex $\ket{h_1,h_2}$ 
in terms of second derivatives of Toda tau-function
$\partial_{t_k} \partial_{t_n} F$ exactly
as it was done in eq.~(\ref{eq:31}). To see how it comes around let us 
integrate over zero-mode  delta-function in the definition of 
two-vertex~(\ref{2vert.N}) and
get rid of one of the zero modes $a_0$. The two-vertex now takes the form
\begin{equation}
 \begin{array}[c]{rl}\ket{V_{12}} =\displaystyle\int dp\,
\exp\Bigg\{
 & 
\displaystyle\frac{1}{2} p^2 (N_{00}^{11} + N_{00}^{22} - N_{00}^{12} - N_{00}^{21}) 
\nonumber \\
&+\displaystyle
\frac{1}{2} p \sum_{n=1}^{\infty} ( N^{11}_{0n} - N^{21}_{0n}) a^1_{-n} 
\nonumber \\
&+\displaystyle
\frac{1}{2} p \sum_{n=1}^{\infty} ( N^{12}_{0n} - N^{22}_{0n}) a^2_{-n} 
\nonumber \\
&+\displaystyle
\frac{1}{2} 
\sum_{I,J=1,2} \sum_{n,m=1}^{\infty} a^I_{-n} N^{IJ}_{nm} a^J_{-m}
\Bigg\}\ket{p}\otimes\ket{-p}
\end{array}
\label{B(N)}
\end{equation}
Here $p$ and $-p$ are eigen-values of operators $a_0^1$ and $a_0^2$
correspondingly. If we identify $a^1_k$ with $a_k$ and $a_k^2$ with $\ba k$
on (see the footnote~\ref{fn:3} \vpageref{fn:3}) and 
equate the above two-vertex to the
boundary state of eq.~(\ref{eq:31}) in accordance with eq.~(\ref{eq:51})
we get the following relations between
Neumann coefficients and second derivatives of Toda tau-function $F$
\begin{eqnarray}
\label{eq:52}
\pfrac{^2 F}{t_0^2}& =& \hphantom{-}N_{00}^{11} + N_{00}^{22} - N_{00}^{12} - N_{00}^{21}\\
\frac{1}{k}\pfrac{^2 F}{t_0 \partial t_k} & =& \hphantom{-} N^{11}_{0k} - N^{21}_{0k} 
\label{f0kh}\\
\frac{1}{k}
\pfrac{^2 F}{{t_0}\partial \bt k} & = &\hphantom{-} N^{22}_{0k} - N^{12}_{0k}
\label{f0barkh}\\
\frac{1}{nk}\pfrac{^2 F}{t_n \partial t_k} & =&  \hphantom{-} N^{11}_{nk}
\label{Id.hol}\\ 
\frac{1}{nk}
\pfrac{^2 F}{{\bar t}_n\partial {\bar t}_k} & = &  \hphantom{-}N^{22}_{nk}
\label{Id.antihol}\\
\frac{1}{nk}\pfrac{^2 F}{t_n\partial {\bar t}_k} & =&-N^{12}_{nk}
\label{Id.mix} 
\end{eqnarray}
This is the identification we were looking for. Several comments are
in order. First of all, 
let us stress one more time that derivatives of the tau-function
$F$ are not independent. They satisfy Hirota identities. 
The corresponding combinations 
of Neumann coefficients should satisfy them too.
In more technical terms it means the following. The second derivatives 
of tau function $\p_{t_n} \partial_{t_k}F$ are related by
Hirota identities eq.~(\ref{hir.eq.pure})-(\ref{eq:122})
to maps $w(z)$ and $\bw(z)$. If one replaces $w(z)$ and $\bw(z)$ with $h_I(z)$
in accordance 
with relation found in the previous section, eq.~(\ref{eq:51}), 
the result should be just a Neumann coefficients in
eq.~(\ref{Id.hol}) - (\ref{Id.mix}). This is indeed the case.
Some subtlety arises in identification of a zero-mode sector due to
different analytical properties of $w(z)$ and $h_I(z)$. We discuss the issue
as well as give all details about the identification in
Appendix \ref{app:2vert-toda}.

From the definition of two-vertex eq.~(\ref{2vert}) one can see that
it is invariant under arbitrary \gl2 transformation of defining conformal maps $h$.
This requires Neumann 
coefficient $N_{km}$ with $k,m>0$ to be invariant under it as well.
As for $N_{0m}$ they are not necessarily 
invariant but must form an invariant combination
after zero-mode $\delta$-function in eq.~(\ref{2vert.N}) being integrated 
over. Not surprisingly, this is exactly the combination one sees in 
eqs.~(\ref{f0kh})-(\ref{f0barkh}).

The identifications~(\ref{eq:52})-(\ref{Id.mix}) were first found in
\cite{Bonora}. There the correct \gl2 invariant combinations
eqs.~(\ref{f0kh})-(\ref{f0barkh}) were guessed. Solving the Hirota identities
gave a value of $\log 16/27$ for $\partial^2_{t_0^2}F$.  Our approach has
several advantages. We were able to derive the identification for all
derivatives of the tau-function including $\partial_{t_0}\partial_{t_k} F$.
We derived the $\partial^2_{t_0^2} F=\log 16/27$ as well.  It is a value of
Neumann coefficients in eq.~(\ref{eq:52}) calculated for a particular choice
of conformal maps, eq.~(\ref{eq:37}). But perhaps more importantly, our
approach makes it very clear that the appearance of the same tau-function in
both objects, the two-vertex and conformally transformed boundary state, is
not a coincidence. This fact will be crucial for us in
Section~\ref{sec:surf-bs}.

\section{Star algebra and dToda tau-function }
\label{sec:3vertex}

Finally we can establish a 
relation between the three-vertex in eq.~(\ref{3vert}) and dToda tau-function.
As before we start with integrating over momentum delta-function in the three-vertex.
The result is eq.~(\ref{3vert}) with $a^3_0 = -a_0 - a'_0$,
$a_0 = a^1_0$ and $a'_0= a^2_0$. We get
\begin{eqnarray}
\ket{V_{123}} &=& \int dpdp'
\exp\Big(
\frac12  \sum_{I,J=1}^3 a_{0}^I a_{0}^J N^{IJ}_{00} +
\nonumber\\
&+&
\sum_{J=1}^3 \sum_{n> 0}
p ( N^{1J}_{0n} - N^{3J}_{0n}) a^J_{-n} +
\sum_{J=1}^3 \sum_{n> 0}
p' ( N^{2J}_{0n} - N^{3J}_{0n}) a^J_{-n} +\\
&+& \frac 12 \sum_{I,J=1}^3 \sum_{m,n>0} a_{-n}^I a_{-m}^J N^{IJ}_{mn} \Big)
\ket{p}_1 \otimes  \ket{p'}_2 \otimes  \ket{-p-p'}_3 \nonumber
\end{eqnarray}
Now we can make the following observation. Since a three-vertex belongs
to the tensor product of Hilbert spaces of three independent open strings
one can multiply it by bra-vacuum belonging to any of the three sectors.
If we choose the vacuum to have zero momentum the result will
be just a two-vertex. For example, let us take a vacuum of the
third string then
\begin{equation}
\label{eq:7.2}
\sprod{0_3}{h_1,h_2,h_3} = \ket{h_1,h_2}
\end{equation}
and similarly for vacua of the first and second string.
All resulting two-vertices can be identified with Toda tau-function
independently. It sets the correspondence between the
three-vertex (in particular the one that defines star-product)
and Toda tau function.

We would like to note that this is what was done in effect in~\cite{Bonora}.
There, one worked with a specific choice of the conformal maps
$h_I(z)$, namely the maps that correspond to Witten's three vertex
eq.~(\ref{eq:37}). In that case there are only
two independent sets of Neumann coefficients, $N^{11}$
and $N^{12}$. The identification was made for 
them only. Therefore, it was actually the identification
with a single two-vertex sector. For such specific choice all other sectors
were equal to this one.

In SFT the choice of the three-vertex is fixed. It means that its defining
conformal maps correspond to tau-function evaluated at some particular values
of times $t_k,{\bar t}_k$ and other values of times never come into play.
There are examples though, when the dependence on conformal maps is important.
For example, the so-called \emph{surface states} are defined with respect to
arbitrary Riemann surfaces (hence their name).  Surface states are conformally
transformed vacua and they are related to tau-function of dispersionless KP
hierarchy~\cite{Bo-Ru}. By varying $\t k$ we are able to change from one
surface state to the other. Thus dKP tau-function describes an infinite 
sequence of surface states, related to each other by conformal
transformations.  The similar statement is true for the dToda tau-function as
we showed in this paper. Indeed, dToda tau-function parameterizes a set of
boundary states on arbitrary curves and thus a set of two-vertices of SFT.
For a given star-product (i.e. for the three-vertex $\ket{V_{123}(h_1,h_2,h_3)}$
with some particular $h_I(z)$'s) and any surface state  $\bra{g}$ defined by a
conformal map $g(z)$ we can define the two-vertex
\begin{equation}
  \label{eq:53}
  \ket{V_{12}(g_1,g_2)}\equiv  \langle{g}\ket{V_{123}} 
\end{equation}
Resulting conformal maps $g_1(z)$ and $g_2(z)$ are now some functionals of
$g(z)$.  Their exact form is determined by three conformal maps $h_I(z)$ of
the three-vertex $\ket{V_{123}}$.  We can repeat this procedure and fuse the
resulting two-vertex with one more surface state, $\ket{S}$. In accordance to
eq.~(\ref{eq:34}) the result will be star-product of $\ket{S}$ with the
initial surface state $\ket{g}$
\begin{eqnarray}
  \label{eq:54}
  \langle S\ket{V_{12}(g)} = \ket{S*g}
\end{eqnarray}
Thus one can view $\ket{g_1,g_2}=\ket{V_{12}(g)}$ as a \emph{multiplication
  operator} on a surface state $\ket{g}$. Hence, the dToda tau-function 
parameterizes the  space of such multiplication operators.

\section{Surface states and boundary states}
\label{sec:surf-bs}

As we have seen in the previous section,  one can associate a
multiplication operator with any state. For a given state $\ket \Sigma$ such
operator $\hat \Sigma$ is defined in the following way
\begin{equation}
  \label{eq:202a}
  \hat \Sigma: \ket X \to \ket{X*\Sigma}\quad \forall\: \ket X
\end{equation}
For a surface state multiplication operator  $\hat \Sigma$ is nothing else 
but a two-vertex $\ket {V_{12}(\Sigma)}$.

As we have shown in Section~\ref{sec:2vertex}, some of the two-vertices can be
interpreted as boundary states. The condition two-vertex should satisfy to be
a boundary state has very simple geometric meaning.  Two conformal maps that
define the two-vertex should map world-sheets of two open strings into
surfaces that can be glued into one world-sheet of a closed string (see
discussion in Section~\ref{sec:2vertex} after eq.~(\ref{eq:51})).  Namely,
world-sheet of the one of the strings should be mapped to the \emph{interior}
of a unit disk, the world-sheet of the other to the \emph{exterior} of the
unit disk. Together they can be combined into the whole complex plane --
world-sheet of the closed string.

We see that although all surface states lead to multiplication operators,
which are some two-vertices, not all of these operators are boundary states
(functions $w(z)$ and $\bar w(z)$ in eq.~(\ref{eq:51}) in general are
\emph{not} complex conjugated).  We come to a very interesting problem of
finding a criteria that the surface state should satisfy to give rise to a
boundary state.  Moreover, the identification with boundary states would
present these surface states in absolutely different light giving them closed
string interpretation.  To address the problem we will use the result of
Generalized Gluing Theorem~\cite{ss}. 

Precise formulation of our problem is the following.  Let the three-vertex be
defined by three conformal maps $h_I(z)$ (say, given by~(\ref{eq:37})), and
the surface state is defined by some conformal map $f(z)$. The contraction of
the tree-vertex and a surface state will give rise to a two-vertex defined by
maps $g_1(z)$ and $g_2(z)$:
\begin{equation}
  \label{eq:55}
  \ket{g_1,g_2}\equiv \la f \ket{h_1,h_2,h_3}
\end{equation}
We want to find the condition which state $\ket f$ should satisfy to give rise
to the two-vertex $\ket {g_1,g_2}$ which can be interpreted as a boundary
state.  Eq.~(\ref{eq:55}) also means, that
\begin{equation}
  \label{eq:56}
  \la f \ket{A*B} = \bra A\otimes \bra B \ket{g_1,g_2},\quad \forall\: A,B
\end{equation}
If we rewrite~(\ref{eq:56}) as
\begin{equation}
  \label{eq:57}
  \sum_{\Phi_r}\la h_1\circ A\, h_2 \circ B \,h_3 \circ \Phi_r\ra\,\la f\circ
  \Phi_r\ra = \la F_1\circ h_1\circ A\, F_1\circ h_2 \circ B\ra
\end{equation}
then Generalized Gluing Theorem~\cite{ss} gives the following answer
\begin{eqnarray}
\label{eq:58}
  g_1 = F_1 \circ h_1(z) ~~~~~ g_2 = F_1 \circ h_2(z)
\end{eqnarray}
where map $F_1$ is defined by some requirements of analyticity that 
we outline below \footnote{for exact formulation see~\cite{ss} } and by
\begin{eqnarray}
  \label{eq:59}
   F_1 \circ h_3(z) = F_2 \circ I \circ f \circ I(z)\equiv  w(z) 
\end{eqnarray}
where $I$ is BPZ inversion $I(z) = -1/z$.  The geometric meaning of
eq.~(\ref{eq:59}) is the following~\cite{ss}. Transformation $h_3(z)$ maps
the interior of the unit disk to interior of some curve ${\cal C}_1$ (wedge
with the angle $2\pi/3$). Correspondingly $f\circ I$ maps the exterior of the
unit disk into the exterior of another curve ${\cal C}_2$.  Let us denote the
compliments of these regions by $D_1$ and $D_2$.  The $F_1$ and $F_2\circ I$
should be such that the images of $D_1$ under $F_1$ and $D_2$ under $F_2\circ
I$ are compliment of each other as well:
\begin{equation}
  \label{eq:60}
 F_2\circ I(D_2) = \IC^1 \backslash  F_1(D_1) 
\end{equation}

This procedure can be visualized as follows. Any vertex can be
represented by a sphere (that is a full complex plane) with holes.  The
boundaries of these holes are images of a unit circle under defining maps $h_I$.
The contraction of two vertexes with maps $h_I$ and $f$ corresponds to gluing
two holes on two spheres together to form a new sphere with other
holes left unchanged. Let $z_1$ be a coordinate on a unit disk to
be mapped by $h_3$ and $z_2$ is coordinate on a unit disk to be mapped by $f$.
The two curves are glued by identification in $z$ plane
\begin{eqnarray} 
  z_1 = -\frac1{z_2}
\end{eqnarray}
To construct a global coordinate $w$ on resulting surface, further maps $F_1$
and $F_2$ are needed.

Before starting the search for the surface states that give rise to the
boundary states let us consider a simple example of so-called \emph{wedge
  state}~\cite{wedgeref}, which will turn out to be useful in the future.

\subsection{Contraction of the wedge states}
\label{sec:wedge}

\emph{Wedge states} are the family of surface states, which form a subalgebra
under star-product. General wedge state $\ket {f_n}$ is defined by means of
conformal map
\begin{equation}
  \label{eq:61}
  f_n(z) = h^{-1}\left(h^{\frac 2n}(z)\right),\quad n>0
\end{equation}
Here $h(z)$ is a famous $SL(2,\IC)$ map, which maps upper half-plane into the
interior of the unit circle and which in particular translates upper half-disk
$\{|z|\le 1, \Im z\ge 0\}$ onto the vertical half-disk $\{|\xi|\le1, \Re \xi
\ge 0\}$:
\begin{equation}
  \label{eq:62}
  h(z) = \frac{1+iz}{1-iz}
\end{equation}
Note, that identity state, vacuum state and sliver state are examples of wedge
states for $n=1,2,\infty$ correspondingly.  The transformation $F_{1,2}$ for a
given wedge state, satisfying properties discussed above
(eqs.~(\ref{eq:59})--(\ref{eq:60})) can be easily found\footnote{Compare with
  similar computations in~\cite{ss}, Section~3.D.} to be
\begin{equation}
  \label{eq:63}
  F_1(u) = u^{\frac 3{n+1}},\quad F_2(v) =
  N\left(e^{i\varphi}v^{\frac{n}{n+1}}\right) 
\end{equation}
(where $N(z)$ is some $SL(2,\IC)$ transformation which will not be
important for us). Eq.~(\ref{eq:63}) means that according to~(\ref{eq:58})
conformal maps for two-vertex, associated with the wedge state~(\ref{eq:61})
via~(\ref{eq:55}) are given by
\begin{equation}
  \label{eq:64}
  \begin{array}[c]{rcl}
  g_1(z) &\equiv & F_1\circ h_1(z) = \Bigl(h^{\frac23}(z)\Bigr)^{\frac3{n+1}}
  = h^{\frac2{n+1}}(z)\\[12pt]
  g_2(z) &\equiv &F_1\circ h_2(z) = \Bigl(e^{\frac{2\pi i}3}
    h^{\frac23}(z)\Bigr)^{\frac3{n+1}} = e^{\frac{2\pi i}{n+1}}
    h^{\frac2{n+1}}(z)
  \end{array}
\end{equation}
Another way to understand eq.~(\ref{eq:64}) is by using the approach
of~\cite{bcft}\footnote{We are indebted to M.~Schnabl for teaching us this
  method.}. Wedge states can be thought of as canonical half disks with added
to them wedges of an angle $\pi(n-1)$~\cite{bcft,schnabl}. Thus, the procedure
of contracting a wedge state with the three-vertex is equivalent to the gluing
two canonical half-disks (corresponding to the transformations $h_{1,2}$ of
three-vertex) to the wedge of the angle $\pi(n-1)$. All together this gives
rise to the cone with the ``angle deficit'' of $\pi(n+1)$. To smooth this
conic singularity into the plane one needs to apply the transformation
$u^{2/(n+1)}$ to the original half-disks. This gives transformations $g_1,g_2$
of~(\ref{eq:64}).

In the next Section we will see that this exercise actually allowed us to find
an example of the boundary state.

\subsection{Example: identity state and orientifold boundary state}
\label{sec:ident}

Let us now come back to the problem stated at the beginning of the
Section~\ref{sec:surf-bs}, the problem of finding
surface states whose multiplication operators are some boundary states.

We can approach it from the other end. Namely, we can first try to find
the smoothing transformations $F_1$ and $F_2$ in eq.~\ref{eq:59}
and only then find the map $f$ that defines the corresponding
surface state.
As we mentioned above (see
Sec.~\ref{sec:2vertex}) in order for two-vertex $\ket {g_1,g_2}$ to be a
boundary state one of the defining maps (say, $g_1(z)$) should map the region
$D$ on which surface state is defined to a unit disk, while $g_2(z)$ would map
the same region $D$ into a compliment of the unit disk.  Represented as a
sphere with holes, such two-vertex will look like a sphere with two holes that
have a common boundary but cover a complimenting regions.
This should be achieved by conformal map $F_1$ applied to two Witten's maps
\begin{eqnarray} 
  \label{eq:65}
  h_1(z) =  h^{2/3}(z)\quad h_2(z) = e^{2\pi i/3} h^{2/3}(z)
\end{eqnarray} 
where $h(z)$ is given by~(\ref{eq:62}). The simplest solution is $F_1(z) =
z^{3/2}$.  Indeed, resulting $F_1\circ h_1$ and $F_1\circ h_2$ will map two
unit disks into the whole complex plane.  We see that such $F_1$ was found for
the case of the wedge state with $n=1$ (eq.~(\ref{eq:63})).  This wedge state is
an identity element of the star-product~\cite{star-alg,schnabl}
\begin{equation}
  \label{eq:66}
  \ket{{\cal I}}*\ket{X} =  \ket{X},~~~~\forall\: X 
\end{equation}
The identity state was originally constructed in~\cite{Ident}.  As a
consequence of being an identity element it has an obvious property -- under
star-product it squares to itself. One can try to interpret it
in the framework of vacuum SFT~\cite{SenBr}.  VSFT equations of
motion for brane ansatz requires the brane solutions to 
square  under star-product to themselves. 
Denote the multiplication operator corresponding to identity state by
$\ket{R}\equiv \ket{V_{12}({\cal I})}$. This $\ket{R}$ is easy to
find~\cite{Ident}.  It should act on any state $\ket X$ like
\begin{equation}
  \label{eq:67}
  \la X \ket R = \ket{X*{\cal I} } = \ket X
\end{equation}
That is $\ket R$ is a \emph{reflector state} -- two-vertex that realizes BPZ
conjugation, i.e. maps any state $\bra X$ into its BPZ conjugated $\ket X$.
From~(\ref{eq:64}) it is obvious that $\ket R$ corresponds to two-vertex with
conformal maps
\begin{equation}
  \label{eq:68}
  g_1(z) = z,\quad g_2(z) = -\frac 1z
\end{equation} 
The explicit form is given by (c.f.  Appendix.~\ref{app:2vert-circ})
\begin{equation}
  \label{eq:69}
  \ket R  = \int dp\,\exp 
\left( -\sum_{n=1}^{\infty} \frac{(-1)^n}{n} a^1_{-n}{a}^2_{-n}  \right)
\ket{p}_1\otimes\ket{-p}_2
\end{equation}
The reflector state is indeed a boundary state. Namely, it is an orientifold
boundary state that evaded an interpretation in VSFT so far, although it is
also a legitimate solution of equations of motion of VSFT. It corresponds to
boundary conditions on a cross-cap~\cite{BS}
\begin{eqnarray}
  \label{eq:70}
  X(\sigma + \pi,\tau) = X(\sigma,\tau) \\
  \partial_\tau X(\sigma + \pi,\tau) = - \partial_\tau X(\sigma,\tau)
\end{eqnarray}
The presence of the orientifold would fit very well in the physical picture of
VSFT. Orientifolds are not dynamical. Unlike D25 branes in bosonic theory they
do not have tachyonic modes and will not decay when tachyon condenses. On the
other hand, orientifolds do couple to closed strings and should be present in
a true closed string vacuum.

Another interesting question comes from the following observation.  Identity
state (as majority of the wedge states) is singular from the point of view of
defining it conformal transformation.  However, contracted with the
three-vertex it became a reflector state, defined by the very simple, \gl2
transformation.  It is possible that this example can help to find a more
general set of surface states that corresponds to some geometrically simple
two-vertices.  Many other states of interest in VSFT correspond to singular
transformations as well.  There is a possibility that their multiplication
operators will be two-vertices built from non-singular transformations as
well.  If true it would make it possible among other things to describe them
in terms of dToda tau-function. We will return to this observation in
Section~\ref{sec:discus}.

\subsection{Involution}
\label{sec:inv}

In the previous section we found one example of a surface state that
leads to an orientifold. It is natural to ask if this solutions is unique
or not. To answer this,
let us come back to the case of three-vertex, contracted with the wedge state
and compare the results~(\ref{eq:64}) with the results of
Section~\ref{sec:2vertex} eq.~(\ref{eq:51}).  As we just saw
(Sec.~\ref{sec:ident}) in case of identity state (i.e.  wedge state with the
$n=1$), the results of contraction is
\begin{equation}
  \label{eq:71}
  g_1(z) = z;\quad g_2(z) = I\circ g_1(z) = -\frac1z
\end{equation}
(with $I(z)$ being BPZ inversion) or \gl2 transformation thereof.

Apparently, eq.~(\ref{eq:71}) is not compatible with the eq.~(\ref{eq:51})
-- the condition for the two-vertex $\ket{g_1,g_2}$ to be a boundary state:
\begin{equation}
  \label{eq:72}
  g_1(z) = w(z)\quad g_2(z) = \frac 1{\bar w(z)}
\end{equation}
The reason for that is that the above condition is a condition
specifically on a Neumann boundary state on a curve. Not just a
boundary state. It is easy to follow the derivation of
eq.~(\ref{eq:72}) to see that it is a consequence of the equation of a unit
circle $z \bz = 1$ (see discussion at the end of the
Section~\vref{sec:2vertex}). Repeating this derivation we can see
that formally eq.~(\ref{eq:71}) is compatible with another curve described by
$\bar z = -1/z$ which can be though as a circle with the \emph{imaginary
  radius} $r=i$. This should be compared with the statement in~\cite{BS},
where it was observed, that orientifold boundary state looks like Neumann
boundary state on the circle of imaginary radius.

This observation suggests the new interpretation to those boundary states
which are \emph{not} the (Neumann) states on the curve. May it have something
to do with the fact that we have a BPZ conjugation in place of the ordinary
(Hermitian) conjugation? If yes, then the question is -- how should we modify
the relation~(\ref{eq:72}) to accommodate for this change.

Recall (see e.g. book~\cite{difr}, Chapter~6.1) that usually in CFT Hermitian
conjugation is defined via $z\to1/\bar z$. That is for the primary field with
the conformal dimensions $(\Delta,\bar \Delta)$ the operation of Hermitian
conjugation is defined as
\begin{equation}
  \label{eq:75}
  [\CO_{\Delta,\bar\Delta}(z, \bar z)]^+ = \bar z^{-2 \Delta} z^{-2 \bar
    \Delta} \CO_{\Delta,\bar\Delta}(1/\bz,1/z) 
\end{equation}
Comparing~(\ref{eq:75}) with~(\ref{eq:72}) we come to the conclusion that
relation between $g_1$ and $g_2$ in~(\ref{eq:72}) is the relation between the
variable and its Hermitian conjugated. Then the natural idea would be to put
$I(z)=-1/z$ in place of Hermitian conjugation eq.~(\ref{eq:72}) to get the
new condition
\begin{equation}
  \label{eq:76}
   \tilde g_1(z) = w(z)\quad \tilde g_2(z) = I\circ w(z) = -\frac1{w(z)}
\end{equation}
Eq.~(\ref{eq:76}) would mean 
\begin{equation}
  \label{eq:77}
  \tilde g_2(z) = I\circ \tilde g_1(z) = -\frac 1{\tilde g_1(z)}
\end{equation}
We will call such two-vertices \emph{BPZ boundary states}\footnote{Another
  suggestion would be to call such states \emph{cross-cap boundary states}} .

From eq.~(\ref{eq:64}) we see that for the two-vertex, obtained from an
arbitrary wedge state~$\ket n$ one has
\begin{equation}
  \label{eq:78}
  g_2(z) = g_1\circ I(z) = g_1(I(z))
\end{equation}
Comparing~(\ref{eq:78}) with~(\ref{eq:77}) one can see, that only those
wedge states would give rise to the BPZ boundary states, which have the
following \emph{necessary} condition
\begin{equation}
  \label{eq:79}
  g_1(z) = I\circ g_1 \circ I(z) 
\end{equation}
It is also immediately clear how to deal with the "general case", when surface
state is \emph{not} the wedge state. Indeed, in that case eq.~(\ref{eq:78})
still 
holds (see~(\ref{eq:58}) together with~(\ref{eq:65})):
\begin{equation}
  \label{eq:80}
  g_1(z) = F_1\circ h^{2/3}(z)\quad g_2(z) = F_1\circ h^{2/3}(I(z))
\end{equation}
(we used the property $h(I(z))=-h(z)$ for~(\ref{eq:62})). Hence,
condition~(\ref{eq:79}) is the criterion for the two-vertices to be
BPZ boundary states.

This gives some new ideas of how to look for boundary states. For example, we
may start by solving the equation~(\ref{eq:79}):
\begin{equation}
  \label{eq:81}
  g(z) g(-1/z) = -1
\end{equation}
Obvious solution of eq.~(\ref{eq:81}) is
\begin{equation}
  \label{eq:82}
  g(z) = z^{2 k +1}, k\in\IN
\end{equation}
Another solution is given by the appropriate regular branch of the
multi-valued function\footnote{This equation was studied in different context
  in~\cite{schnabl.state}.}
\begin{equation}
  \label{eq:83}
  g(z) = z^{\frac1{2 k +1}}, k\in\IN,\quad 
  \mbox{such branch that}~g(-1) = -1 
\end{equation}
The solution in~(\ref{eq:83}) should be 
compared with eq.~(\ref{eq:64}) to see that
it describes wedge state~$\ket{4 k + 1}$ $k\ge0$. Thus, such family of wedge
states describes BPZ boundary states. Note, that multiplication rule of
wedge states is~\cite{schnabl}
\begin{equation}
  \label{eq:84}
  \ket r * \ket s  = \ket{r+s-1}
\end{equation}
and as a result these surface states form a subalgebra:
\begin{equation}
  \label{eq:85}
  \ket {4 k+1} * \ket{4 m+1}  = \ket{4(k+m)+1}
\end{equation}

On the other hand, solution~(\ref{eq:82}) means that function $F_1(z)$
of~(\ref{eq:58}) is such that
\begin{equation}
  \label{eq:86}
  z^{2 k +1} = F_1(h^{2/3}(z)) \Rightarrow F_1(u) =
  \Bigl(h^{-1}(u^{3/2})\Bigr)^{2 k +1}
\end{equation}
Geometric meaning of this function is not clear to us at the moment, but it is obvious, that using it and eq.~(\ref{eq:59}) one
can very simply obtain the function $f(z)$ (in r.h.s. of~(\ref{eq:59})), and
thus the corresponding surface state.

Whether there are more solutions of~(\ref{eq:81}) is not clear to us at the
moment. Note, that they are \emph{necessary} but not \emph{sufficient}
conditions! For each of them, the procedure of finding $f(z)$ via
eq.~(\ref{eq:59}) should still be realized. It would be very interesting to
check these two solution and see the class of $f(z)$, which correspond to
them.

\subsection{Discussion}
\label{sec:discus}

As we mentioned in the introduction the D-branes as a solitons in open
strings arise in the context of Vacuum SFT~\cite{SenBr}.
It would be interesting to build the multiplication operator~(\ref{eq:202a})
for projectors, corresponding to the D-branes in VSFT, to see whether they
correspond to boundary states, In first place we would like to do it for the
sliver state. This is a state conjectured to correspond to
space-filling D-brane in VSFT. It would be interesting
to see explicitly that it is a Neumann boundary state (that of a
space-filling D-brane) of some curve and find out the geometric meaning of
this curve.  This was partially fulfilled in the Section~\ref{sec:wedge},
eq.~(\ref{eq:64}). However it is hard to analyze it fully in case of sliver,
i.e. the singular limit $n\to\infty$ in eq.~(\ref{eq:61}).  Right now we
would just like to stress one existing connection.  Sliver state is a surface
state that corresponds to the conformal map on infinitely-sheeted
logarithm-like Riemann surface~\cite{star-alg}.  On the other hand, the
two-vertex~(\ref{eq:49}), corresponding to the Neumann boundary state, has
logarithmic multi-valuedness as well. Indeed, recall
(Section~\ref{sec:cft-vac}) that the scalar field we considered had a term
$\hb0 \log w +\bb0 \log {\bar w}$.  In order for it to be single-valued on the
complex plane $w$, $\hb0$ should be equal to $\bb0$ (compare
footnote~\ref{fn:1} \vpageref{fn:1}).  As we had mentioned already in
Section~\ref{sec:bcft} we did not impose this condition, but instead required
Neumann boundary conditions $\hb0 = - \bb0$.  As a result the scalar CFT and
Neumann boundary state should be defined on multi-sheet logarithm Riemann
surface as well.  It will be very interesting to see how this Riemann surface
and Riemann surface of a sliver state are related.

We would like also to note, that for the subalgebra of states, described
in~\ref{sec:inv} (namely, states~(\ref{eq:83})) we can formally take limit
$k\to\infty$ and thus it may be that sliver shares some of their properties.
We leave this question to the future investigation.

To see, whether the conjectured relation between projectors and boundary
states is true, one would have to perform several checks. One of them would be
to compute one-point function of closed string vertex operators \emph{in the
  background of given solution of SFT}. For that one would have to act as
follows. First, compute the two-vertex $\ket{V_{12}(\Sigma)}$, corresponding
to the given surface state $\ket \Sigma$. For example this state can be
projector, conjectured to correspond to a D-brane. Then, interpret this vertex
as a boundary state $\bket\Sigma$ (as discussed in details in
Section~\ref{sec:2vertex}).  Finally, compute the one-point function of any
closed string vertex operator $V_{closed}(z,\bar z)$:
\begin{equation}
  \label{eq:98}
  \bra 0 V_{closed}(z,\bar z)\bket\Sigma \stackrel{?}=\langle
  V\rangle_{disk,\; Dirichlet} 
\end{equation}
where correlator  $\langle V_{closed}\rangle$ in
the r.h.s. of~(\ref{eq:98}) is computed on the disk with appropriate
(Dirichlet) boundary conditions.

Another interesting check would be to calculate ratios of tensions
corresponding to branes of different dimensions. In VSFT the tension is just a
norm of a projector state $\ket{\Sigma}$.  On the other hand the tension of a
brane is extracted from the amplitude of closed string exchange between the
branes
\begin{equation}
  \label{eq:99}
  \bbra{V_{12}(\Sigma)} 
  \exp^{-(L_0^1 + L^2_0)\pi/Y}
  \bket{V_{12}(\Sigma)} = \\
  \bra{\Sigma} \bra{V_{1^*23}} 
  \exp^{-(L_0^1 + {L}_0^2)\pi/Y}
  \ket{V_{231^*}} \ket{\Sigma}
\end{equation}
where star in $\bra{V_{1^*23}}$ and $\ket{V_{231^*}}$ means that corresponding
state is BPZ conjugated. The parameter $Y$ is a separation between branes.  We
do not that hope this norm will be equal to the norm of $\ket \Sigma$ itself.
But it is possible that the ratio of tension of different branes will stay the
same.

The interpretation of two-vertices as boundary states is useful 
realization of a star-product in closed SFT.
To wit, it was suggested recently in~\cite{closedSFT} 
to describe D-branes as boundary states in closed SFT. It
was also shown there that by taking so called HIKKO formulation of closed
SFT~\cite{hikko}  boundary states $\bket B$ corresponding to
various D-brane solutions indeed obey the equation
\begin{equation}
  \label{eq:137}
  \bket B \star \bket B = \bket B
\end{equation}
That is, boundary states formally satisfy the same equations
that are imposed on brane solutions in VSFT. The star-products
in both theories are very different, of course.
This prompts the possibility of a more general correspondence.
Let us define the star-product of boundary states as
usual Witten's star-product of corresponding surface states.
Namely, if open string surface states $\ket {f_1}$, $\ket {f_2}$ and 
$\ket{f_3}$ give rise in a way suggested in our paper to boundary states
and
\begin{equation}
  \label{eq:140}
  \ket{f_1}*\ket {f_2}=\ket {f_3}
\end{equation}
let us define the result of a star-product of
$\bket {B_1} \star \bket {B_2 }$ as a state $\bket {B_3}$ that 
corresponds to $\ket{ f_3}$.
Note, that in this case eq.~(\ref{eq:137})
becomes the natural consequence of the hypothesis of~\cite{SenBr} that surface
states, corresponding to D-branes square to themselves under Witten's star
product~\cite{wittenSFT}. We do not know
if such star-product is indeed a HIKKO star product or if it is 
a consistent star product at all.
But if correct, this will be a very interesting observation.
At this moment, though, this is just a conjecture and we leave it for
future works. Let us just mention immediate consequence of this conjecture:

\begin{itemize}
\item[--] boundary state~(\ref{eq:69}), corresponding to orientifold, should
  serve as an \emph{identity element} of the $\star$-algebra of closed SFT.
\end{itemize}

One of the main points of the paper is introduction of full tau-function
dynamics in SFT. The SFT vertices we considered depended on a family
of conformal maps. In terms of tau-function times $t_k$ it means
that Neumann coefficients of corresponding vertices become functions
of $t_k$ as well. This result opens the whole field of problems
that can be pursued. 
\begin{itemize}
\item 
Since surface states are just one-vertices and surface states form a
sub-algebra under star product it should be possible to re-write star
multiplication analytically in terms of $t_k$. This would allow us ,
in particular, to use the well-known integrable reductions to identify the
finite-dimensional sub-algebras of star-product.  
\item
In~\cite{wwdvv} it was
shown that associativity (WDVV) equations~\cite{wdvv} are solved by
tau-function as a direct consequences of Hirota identities. The integrable
structure of SFT, identified in this paper, thus strongly suggests the
presence of such associativity algebra. Technically, it will involve
the derivatives of Neumann coefficient w.r.t.\ 
$t_k$, i.e.\ the third derivatives of tau-function.
The identification of the chiral rings
of associative operators in SFT is just one of the numerous possibilities
opened by integrability.
\end{itemize}

\section*{Acknowledgements}

We would like to acknowledge useful communications with J.~Ambj\o{}rn,
J.~Harvey, A.~Losev, N.~Nekrasov, V.~Schomerus, B.~Zwiebach. We are very
grateful to Martin Schnabl for many useful communications during the
preparation stage of this paper as well as for reading the draft and providing
many useful comments.  A.B.  and O.R. would like to acknowledge the warm
hospitality of IH\'ES where part of this work was done.  O.R. would also like
to thank NBI.  The work of BK is supported by
German-Israeli-Foundation, GIF grant I-645-130.14/1999

\appendix

\renewcommand{\thesubsection}{\Alph{section}\arabic{subsection}}

\section{Toda Lattice Hierarchy and its dispersionless limit}
\label{app:toda}

In this Appendix we will review some facts about Toda Lattice hierarchy.

\subsection{Toda Lattice Equation}
\label{sec:toda-latt-equat}

Equation of (dispersionful) Toda Lattice is given by
\begin{eqnarray}
\label{toda}
\partial_{\t1}
\partial_{\bt1}\phi_n=e^{\phi_n-\phi_{n-1}}-e^{\phi_{n+1}-\phi_{n}} 
\end{eqnarray}
It is convenient to introduce a new variable $\t0$, function $\phi(\t0)$ and
\emph{lattice spacing} $\hbar$ such that 
\begin{equation}
  \label{eq:100}
  \phi_n = \phi(\t0),\; \phi_{n\pm1} = \phi(\t0\pm\hbar),\;\mbox{etc.}
\end{equation}
Eq.~(\ref{toda}) is known to be \emph{integrable} (see
e.g.~\cite{ueno-tak,tak-tak,Adler} and refs. therein).
In particular it  has the so called \emph{Zakharov-Shabat} or \emph{zero
  curvature} representation
\begin{equation}
  \label{eq:101}
  \p_{\t1} \bar H_1 - \p_{\bt1} H_1 + [\bar H_1,\,H_1] = 0
\end{equation}
where infinite matrices $H_1$ and $\bar H_1$ are
defined~\cite{mikhailov} as 
\begin{equation}
  \label{eq:102}
  H_1 = \left (
  \begin{array}[c]{cc|ccccc}
    \ddots&\ddots & \vdots & \vdots & \vdots &
    \vdots & \vdots \\ 
    \cdots &   \p_{\bt1}\phi(\t0-\hbar)& 1 & 0 & \cdots\\
\hline
    \cdots &  0 &\p_{\bt1}\phi(\t0)& 1&\cdots\\
    \cdots &  0 & 0 & \p_{\bt1}\phi(\t0+\hbar)& \ddots\\
    \cdots &  \vdots & \vdots & \vdots & \ddots 
  \end{array}\right)
\end{equation}
and
\begin{equation}
  \label{eq:103}
  \bar H_1 = \left (
  \begin{array}[c]{cc|ccccc}
    \ddots & \vdots & \vdots & \vdots & \vdots \\
    \cdots & 0 & 0 & 0 & \cdots\\
    \hline
    \cdots &  e^{\phi(\t0)-\phi(\t0-\hbar)} & 0 & 0 & \cdots\\
    \cdots & 0 &  e^{\phi(\t0+\hbar)-\phi(\t0)} & 0 & \cdots\\
    \cdots &  \vdots & \vdots & \vdots & \ddots 
  \end{array}\right)
\end{equation}
Toda Lattice equation can also be written in the form which introduces a
\emph{tau-function} $\tau(\t0,\t1,\tb1)$. Namely, if one defines
\begin{equation}
  \label{eq:104}
  \phi(\t0)= \log\frac{\tau(\t0+\hbar)}{\tau(\t0)},\;
  e^{\phi(\t0)-\phi(\t0+\hbar)} = \frac{\tau(\t0+\hbar)\tau(\t0-\hbar)}{\tau(\t0)^2}
\end{equation}
then equation~(\ref{toda}) can be rewritten in the \emph{Hirota
form}
\begin{equation}
  \label{eq:105}
  \frac 12 D_{\t1}D_{\tb1} \tau(\t0)\tau(\t0) + \tau(\t0+\hbar)\tau(\t0-\hbar)=0
\end{equation}
where \emph{Hirota derivative} $D_{x}D_{y} f(x,y)f(x,y)$ is defined via
\begin{equation}
  \label{eq:106}
   D_{x}D_{y} f(x,y)f(x,y) \equiv 2\left( \frac{\p^2 f(x,y)}{\p x \p y}
   f(x,y)-\frac{\p f(x,y)}{\p x}\frac{\p f(x,y)}{\p y}\right)
\end{equation}
We will see below that the tau-function 
will play an important role in describing the hierarchy.

\subsection{Toda Lattice Hierarchy}
\label{sec:toda-latt-hier}

Eq.~(\ref{eq:101}) is just a first equation of an infinite series of zero
curvature equations that described the whole \emph{Toda Lattice
  Hierarchy}. This hierarchy can be represented in the following form. Let us
define two \emph{Lax operators}
\begin{equation}
  \label{eq:107}
  L = r(\t0)e^{\hbar \p_{\t0}} + \sum_{k=0}^\infty u_k(\t0)e^{-k\hbar
    \p_{\t0}}
\end{equation}
and
\begin{equation}
  \label{eq:108}
  \bar L = e^{-\hbar \p_{\t0}}r(\t0) + \sum_{k=0}^\infty e^{ k\hbar \p_{\t0}}\bar u_k(\t0)
\end{equation}
Then one can build (double infinite) series of \emph{Lax-Sato equations}
\begin{eqnarray}
  \label{eq:109}
  \frac{\p L}{\p \t k} & =& [H_k,\,L]\\
  \label{eq:110}
  \frac{\p L}{\p \bt k}& =& [\bar H_k,\,L]
\end{eqnarray}
defining the evolution with respect to all times $\t k$, $\tb
k$. Here\footnote{
  Sometimes eq.~(\ref{eq:111}) is written in the form $H_k =
  \left(L^k\right)_+$, etc. This is the gauge choice, depending on the form of
  the Lax operators~(\ref{eq:107})}
\begin{equation}
  \label{eq:111}
  H_k = \left(L^k\right)_++\frac12\left(L^k\right)_0;\quad \bar H_k =
  \left(\bar L^k\right)_-+\frac12\left(\bar L^k\right)_0 
\end{equation}
Compatibility of the system of equations~(\ref{eq:109}--\ref{eq:110}) leads to
Zakharov-Shabat equations
\begin{eqnarray}
  \label{eq:112}
  \frac{\p H_n}{\p \t k} -\frac{\p H_k}{\p \t n} + [H_n,\,H_k] &=& 0\\
  \label{eq:113}
  \frac{\p \bar H_n}{\p \t k} -\frac{\p H_k}{\p \bt n} + [\bar H_n,\,H_k] &=& 0
\end{eqnarray}
(there is also an 
equation analogous to~(\ref{eq:112}) for Hamiltonians $\bar H_k$
and variables $\bt n$). Eq.~(\ref{eq:101}) was the first in
series~(\ref{eq:112}). The identification between eq.~(\ref{eq:101})
and~(\ref{eq:112}) is established by
\begin{equation}
  \label{eq:114}
  r^2(\t0) = e^{\phi(\t0+\hbar)-\phi(\t0)};\quad u_0(\t0) =
                   \frac{\p \phi(\t0)}{\p \t1};\quad \mbox{etc}
\end{equation}

\subsection{Tau-function of Toda Lattice Hierarchy}
\label{sec:tau-function-toda}

There are many (to some extent) equivalent ways to describe integrable
hierarchies. In various applications various of them are useful. For example,
above we have shown two such formalisms: ``Lax formalism'', described by
eqs.~(\ref{eq:107})--(\ref{eq:111}) and Zakharov-Shabat formalism, described
by eqs.~(\ref{eq:112})--(\ref{eq:113}). Below, we are going to introduce one
more way to describe the hierarchy, which is going to be extremely useful for
us in what follows - the so called \emph{tau-function approach}.

One can introduce a tau-function for the whole Toda Lattice Hierarchy in the
following way. Consider eqs.~(\ref{eq:109}), (\ref{eq:110}) as a compatibility
condition for the auxiliary linear problem for function
$\Psi(z|t)\equiv\Psi(z|\t0,\t k,\bt k)$
\begin{equation}
  \label{eq:115}
  L \Psi(z|t) = z \Psi(z|t),\;
  \pfrac{\Psi(z|t)}{\t k} = H_k \Psi(z|t),\;  k>0 
\end{equation}
Function $\Psi(z|t)$ are called \emph{Backer-Akhiezer functions}.

The solution of all the equations can
be incorporated into a function of infinite variables $\tau(\t0,\t k, \bt
k)$ such that
\begin{equation}
  \label{eq:116}
  \Psi(z|t) = \frac{\tau\left(\t0;\vt - \h\left[z^{-1}\right];\bar
      \vt\right)}{\tau(\t0;\vt;\bar \vt)} e^{\frac{\xi(z)}{\h}} z^{\frac{\t0}{\h}} 
\end{equation}
where $\xi(z)\equiv \sum_{k=1}^\infty \t k z^k$ and notation
$\left[z^{-1}\right]$ means
\begin{equation}
  \label{eq:117}
  f\left(\vt - \left[z^{-1}\right]\right) =  f\left(\t 1 - \frac 1{z},\t2 -
    \frac 1{2 z^2},\dots\right)
\end{equation}
Similar equations can be written for the conjugated function $\Psi^*(z)$
\begin{equation}
  \label{eq:118}
    \Psi^*(z|\t0,\t k,\bt k) \equiv \Psi^*(z|t)=
  \frac{\tau\left(\t0;\vt + \h\left[z^{-1}\right];\bar
      \vt\right)}{\tau(\t0;\vt;\bar \vt)} e^{-\frac{\xi(z)}\h}
  z^{-\frac{\t0}{\h}}  
\end{equation}

In terms of Backer-Akhiezer functions the whole Toda Lattice hierarchy can be compactly
encoded in the form
\begin{equation}
\label{ba}
\oint_\infty \frac{d z}{2\pi i} \Psi(z|\t0,\t k,\bt k) \Psi^*(z|\t0',\t k', \bt k') = 0
\end{equation}
That is, one can show that condition~(\ref{ba}) is equivalent to the
system of equations~(\ref{eq:107})--(\ref{eq:111}) and thus equivalently describes Toda
Lattice hierarchy.

Eq.~(\ref{ba}) can be re-expressed in terms of infinite set of differential
equations on tau-function, called \emph{Hirota equations}\footnote{
  The
  following result for Toda Lattice hierarchy Hirota identities 
  was shown to us by A.Zabrodin. We are  grateful to him for sharing with us
  this information.}. 
This form of Hirota  identities is similar to the one appearing 
in the case of KP (see~\cite{kodama}). For a general form, 
see also~\cite{ueno-tak}.

Introduce the operator
\begin{equation}
\label{eq:119}
D(z) = \sum_{k=1}^{\infty} \frac{z^{-k}}{k}\de_{t_k}
\end{equation}
Then eq.~(\ref{ba}) can be rewritten in the following form
\begin{equation}
  \label{eq:120}
  \begin{array}[c]{l}
       z\left(e^{\h(\p_{\t0} -D(z))}\tau\right)\left(e^{-\h D(\zeta)}\tau\right) -
    \zeta\left(e^{\h(\p_{\t0} -D(\zeta))}\tau\right)\left(e^{-\h D(z)}\tau\right)\\\\
    = (z-\zeta)\left(e^{-\h (D(z)+ D(\zeta))}\tau\right)\left(e^{-\h
      \p_{\t0}}\tau\right)
\end{array}
\end{equation}
\begin{equation}
  \label{eq:121}
  \begin{array}[c]{l}
    \left(e^{-\h D(z)}\tau\right)\left(e^{-\h \bar D(\bar \zeta)}\tau\right) -
    \tau \left(e^{\h (\bar D(\bar \zeta) -D(z))}\tau\right) \\ \\
    = \dfrac 1{\strut z\bar \zeta}\left(e^{\h(\p_{\t0} +
        D(z))}\tau\right)\left(e^{\h(\p_{\t0} +\bar D(\bar \zeta))}\tau\right)
  \end{array}
\end{equation}
There are also other Hirota equations, which we do not provide here.

\subsection{Dispersionless limit of Toda Lattice Hierarchy}
\label{sec:disp-limit-toda}

Many integrable hierarchies (Toda Lattice being one of them) admit the so
called \emph{dispersionless limit}. New hierarchies are also integrable,
according to the general theory of Witham hierarchies and hydro-dynamical
brackets~\cite{kri,dn}.  These hierarchies also appear in many applications.

In the case of Toda, to obtain the dispersionless limit one takes formal limit
$\h\to0$ in eq.~(\ref{toda}),~in identification~(\ref{eq:100}),~(\ref{eq:107}),
etc.  In this limit, for example, the equation~(\ref{toda}) takes the form (see
e.g.~\cite{tak-tak})
\begin{eqnarray}
\label{dToda}
\partial _{\t1} \partial _{\bar
  \t1}\phi(\t0)=\p_{\t0}e^{\partial_{t_0}\phi(\t0)} 
\end{eqnarray}
This equation is also integrable and is a part of dispersionless Toda lattice
hierarchy. Again, all equations of this hierarchy can be
encoded in the Hirota identities, which can be obtained by taking the same
limit $\h\to0$ in~(\ref{eq:121}) (writing
$\tau_\h=\exp(F/\h^2+\CO(1))$, etc.)
\begin{eqnarray}
\label{dHirota}
  (z-\zeta)e^{D(z)D(\zeta)F} = ze^{-\partial_{\t0}D(z)F} - \zeta e^{-\partial_{\t0}D(\zeta)F}\
\end{eqnarray}
\begin{eqnarray}
\label{dHirota-mixed}
1-\exp(-D(z)\bar D(\bar \zeta)F) =\frac1{z\bar\zeta}\exp(\p_{\t0}(\p_{\t0} +
    D(z) + \bar D(\bar \zeta))F), 
\end{eqnarray}
where $D(z)$ is defined by~(\ref{eq:119}).  There is also Hirota identity in
purely anti-holomorphic sector, it is identical to the holomorphic one~(\ref{dHirota}).

The map $w(z)$ understood as a function of $t_k$ is given by
\begin{eqnarray}
  \log \frac{w(z)}{z/r} = - \de_{t_0} D(z) F({\bf t}) 
  \label{hir.0}
\end{eqnarray}
and $\log r^2 = \de^2_{t_0}F({\bf t})$.
This allows to rewrite Hirota equations~(\ref{dHirota}),~(\ref{dHirota-mixed})
\begin{eqnarray}
  D(z)D(\zeta)F({\bf t}) - \frac{1}{2} \de^2_{t_0}F({\bf t}) = 
  \log \frac{w(z) - w(\zeta)}{ z - \zeta } \label{hir.eq.pure}
\end{eqnarray}
and
\begin{equation}
  \label{eq:122}
  D(z)\bar D(\zeta)F = -\log\left(1 - \frac1{w(z)\bar w(\zeta)}\right)
\end{equation}
Coupled with eq.~(\ref{hir.0}) one can see that second derivatives
$\de_{t_k}\de_{t_n} F$ for $k,n>0$ as well as $\de_{\bt k}\de_{\bt n} F$ are completely determined by 
$\de_{t_0}\de_{t_k} F$.

By expressing from~(\ref{dHirota}) derivatives $\p_{\t0}\p_{\t k}F$ 
via $\p_{\t1}\p_{\t k}F$, 
Hirota eqs.~(\ref{dHirota})(\ref{hir.eq.pure}) can also be rewritten
in the \emph{pure holomorphic form}, which is also called \emph{Hirota
  equations of dispersionless KP hierarchy}~\cite{kodama}
\begin{equation}
  \label{eq:123}
  \exp(D(z)D(\zeta)F) = 1 - \frac{D(z)\p_{\t1}F-D(\zeta)\p_{\t1}F}{z-\zeta}
\end{equation}

\section{Two-vertex corresponding to Neumann boundary state on a circle}
\label{app:2vert-circ}

In this Appendix we want to make an explicit calculation for the
two-vertex corresponding to conformal transformations, $h_1(z) = z$ and
$h_2(z) = 1/z$. The Neumann coefficients are
\begin{eqnarray}
N^{12}_{nm} &=& N^{21}_{mn} = \oint\oint \frac{dzdw}{(2\pi i)^2} \frac{z^{n}w^{m}}{nm}
\frac{-1}{z^2} \frac{1}{\big(1/z - w \big)^2} 
= -\frac{1}{n} \delta_{n,m} \\
N^{11}_{nm} &=& N^{22}_{nm} = \oint\oint \frac{dzdw}{(2\pi i)^2} 
\frac{z^{n}w^{m}}{nm}
\frac{1}{z^2}\frac{1}{w^2} \frac{1}{\big(1/z - 1/w \big)^2}
= -\frac{1}{m} \delta_{-n,m}
\end{eqnarray}
In the definition of the 
vertex state eq.~(\ref{2vert.N}) the sum is over positive
$n$, thus only $N^{12}$ contributes. Others are contracted with lowering 
operators and they annihilate the vacuum. Now we can bring this together
to get 
\begin{eqnarray}
\label{ap:1}
\ket{V_{12}} =  
\int dp \exp{ \left\{ - \sum_{n>0} \frac{a^1_{-n} a^2_{-n}}{n} \right\}}  \ket{p}_1\otimes \ket{-p}_2
\end{eqnarray}
We assumed that zero-mode Neumann coefficients that stand in front 
of $p$ and $p^2$ terms are zero (see eq.(\ref{B(N)})). In dealing with
these Neumann
coefficients $N^{IJ}_{0k},\:k\ge0$ one has to manipulate with formal objects
like $h_I(\infty)$, $[h'](\infty)$. Let us demonstrate  how to do it.
We make use of the following observation.
The choice $h_1(z) = z$ and 
$h_2(z) = 1/z$ is not unique. One can take any \gl2 transformations
of $h_I(z)$. It is easy to see that Neumann coefficients are left unchanged.
Thus a general choice for conformal transformations that leads to
Neumann boundary state 
is given by~(see the footnote~\ref{fn:5}, p.~\pageref{fn:5})
\begin{equation}
  \label{eq:124}
  h_1(z) = \frac{a z + b}{c z + d},\quad    h_2(z) = \frac{(1/z) a +
    b}{(1/z) c + d}
\end{equation}
If any of the $h$'s are not regular at infinity
we can build new $h$'s which are regular by applying any \gl2 transformation. 
In this way we can ``regularize'' $h_I(\infty)$.
Let us take this \gl2 transformation to be small.
For example, to deal with current $h_{1,2}(z)$ we choose the following \gl2 matrix
\begin{equation}
  \label{eq:125}
  \left(
    \begin{array}[c]{cc}
      1 & 0\\
      \varepsilon & 1
    \end{array}\right)
\end{equation}
Then
\begin{equation}
\label{fn:6}
    h_1(z) = \frac z {1 + \varepsilon z} \quad
    h_2(z) = \frac 1 {z + \varepsilon} 
\end{equation}
Now $h_{1,2}(\infty)$ are well defined. We can make the
calculations 
and take the limit $\varepsilon \rightarrow 0$ at the end.
As an example
\bea
N^{11}_{0n} = \frac 1n \oint\frac{dz}{2\pi i} z^n 
\de_z \log\LB \frac 1\varepsilon - \frac z {1 + \varepsilon z} \RB =
0~~~\forall\: n>0
\eea
The coefficient of the $p^2$ term is\footnote{
  Recall, that $[h'](z)$ is defined as $\displaystyle [h'](z) \equiv
  - \frac{d h(z)}{d 1/z}$ 
}%
\bea
\log\left\{ - \frac{[h'_1](\infty)[h'_2](\infty)}{\LB
h_1(\infty)-h_2(\infty) \RB^2}\right\} = 
\log \frac{\varepsilon^2}{\varepsilon^2} = 0
\eea
Thus the eq.~(\ref{ap:1})  is indeed Neumann boundary state on a circle as
one can see from eq.~(\ref{eq:24}).

\section{Conformal transformations of a two-vertex state}
\label{app:2vert-conf}

Below, we give a proof that a two-vertex defined by arbitrary
conformal maps corresponds to a Neumann boundary state 
defined on conformally transformed
unit circle. Consider a curve $\CC$ which is mapped to a unit
circle by conformal transformations $w(z)$.
Let us define the action of conformal transformations on operators and states.
Under transformation $g(z)$
\begin{eqnarray}
  g(V) \equiv U_g V U^{-1}_g \label{conf.op2}\\
  g \ket{\psi} \equiv  U_g \ket{\psi} \\
  {\rm So~that~~~} g\Big(V \ket{\psi}\Big) = 
  g(V) g\ket{\psi}
\end{eqnarray}

Here $U_g$ are the elements of Virasoro group
given by
\begin{equation}
  \label{eq:126}
  U_g = \exp(\sum_{n} v_n L_n)
\end{equation}
where $v_n$'s are harmonics of $v(z) = \sum
v_n z^{n+1}$ and field $v(z)$ is defined by  equation
\begin{equation}
  \label{eq:127}
  e^{v(z)\p_z} z = g(z)
\end{equation}
If the 
conformal transformation $g(z)$ is regular at infinity the expansion of
the field $v(z)$ has only modes with $n\le1$. As a result we have
\begin{equation}
  \label{eq:128}
  \bra0 U_g=\bra0 U_g^{-1}=\bra0
\end{equation}
Again, denote the two-vertex~(\ref{2vert}) defined by maps $h_1$ and $h_2$ via
$\ket{h_1,h_2}$.
Therefore what we showed in Appendix~\ref{app:2vert-circ} can be written as
$\ket{z,{z}^{-1}} = \bket{N}_{circle}$, 
with $\bket{N}_{circle}$ being Neumann boundary state
on a circle. 
The task is now to find which conformal transformations 
$w(z)$ and ${\bar  w}(\bar z)$ 
correspond to such $h_1(z)$ and $h_2(z)$ in
definition of two-vertex that the resulting two-vertex 
is conformally transformed Neumann boundary state
\begin{eqnarray}
  \ket{h_1,h_2} = U_{w}^{-1} \bar U_{{\bar w}}^{-1} \bket{N}_{circle}
  \label{def.h.w} 
\end{eqnarray}
where $U_w$ acts on a holomorphic part of $\bket{N}$ and 
${\bar U}_{\bar w}$ acts on anti-holomorphic part. 
$U^{-1}_w$ comes from the fact,
that $a_k = U_w b_k U^{-1}_w$ and $b_k =  U^{-1}_w a_k U_w$.

To do this, 
let us calculate the two-vertex corresponding to composite conformal
transformation $f_1 \circ w$ and $f_2\circ \bar w$. We can do it in two
steps. We can either 
apply the whole $f \circ w$ at once or take $f$ first and $w$ later.
For any vertex operators $V$ one can write using definition of two-vertex
eq.~(\ref{2vert})
\begin{equation}
\renewcommand{\arraystretch}{1.2}
  \begin{array}[c]{rcl}
    \bra{0}_1\otimes\bra{0}_2 V_1(a^1) V_2(a^2)
    \ket{f_1\circ w_1,f_2\circ w_2}
    &=& \\
    \bra{0} f_1 \circ w_1 (V_1)  f_2 \circ w_2 (V_2) \ket{0}
    &=& \\
    \bra{0}_1\otimes\bra{0}_2 w_1(V_1) w_2(V_2)
    \ket{f_1,f_2} & &
\end{array}
\label{vert.h.and.f}
\end{equation}
Now, we can use the explicit form $f(V)$ from eq.~(\ref{conf.op2}) to get
\begin{equation}
\renewcommand{\arraystretch}{1.5}
  \begin{array}[c]{rcl}
  \bra{0}_1\otimes\bra{0}_2 w_1(V_1)  w_2(V_2) \ket{f_1,f_2}
  =\nonumber \\ 
  \bra{0}_1\otimes\bra{0}_2 U_{w_1} V_1(a_1) U^{-1}_{w_1} U_{w_2} V_2(a_2)
  U^{-1}_{w_2} 
  \ket{f_1,f_2}  = \nonumber \\
  \bra{0}_1 \otimes \bra{0}_2
  V_1(a_1) V_2(a_2) U^{-1}_{w_1} U^{-1}_{w_2}\ket{f_1,f_2} 
\end{array}
\label{vert.h}
\end{equation}
In the last line we assumed that $w_1$ and $w_2$ leave the vacuum
states invariant eq.~(\ref{eq:128}). 
Comparing eq.~(\ref{vert.h}) and eq.~(\ref{vert.h.and.f}) we see that
for any $V_1$ and $V_2$
\begin{eqnarray}
&&  \bra{0}_1\otimes\bra{0}_2  V_1(a^1) V_2(a^2)
  \ket{f_1\circ w_1,f_2 \circ w_2} = \\
&&  \bra{0}_1 \otimes \bra{0}_2
  V_1(a^1) V_2(a^2) U^{-1}_{w_1} U^{-1}_{w_2} \ket{f_1,f_2}
\end{eqnarray}
Thus we can say that corresponding states are equal
\begin{eqnarray}
\ket{f_1\circ w_1,f_2 \circ w_2} =
U^{-1}_{w_1} U^{-1}_{w_2} \ket{f_1,f_2}
\end{eqnarray}
For the fixed choice
of $f_1(z)={z}$ and $f_2(z)=\frac{1} z$, 
$\ket{z,\frac1z}$ is just a Neumann boundary state $\bket{N}_{circle}$ 
on a unit
circle, defined by eq.~(\ref{eq:24}) (see also
Appendix~\ref{app:2vert-circ}).
If $w_1(z) = w(z)$ and  $w_2(z) = {\bar w}(z)$ then
\begin{eqnarray}
  \label{eq:129}
  \ket{w,\frac1{\bar w}} = U_{w}^{-1} \bar U_{{\bar
      w}}^{-1} \bket{N}_{circle} 
\end{eqnarray}
It should be compared to eq.~(\ref{def.h.w}).  Thus we achieved the
result~(\ref{eq:51}) --- the two-vertex with conformal transformations
\begin{equation}
  h_1=w(z),\quad h_2(z)=\frac1{{\bar w}(z)}
  \label{eq:130}
\end{equation}
is a Neumann boundary state $\bket{N}_\CC$ on a curve~$\CC$.

Now we come back to the question, raised in the
footnote~\ref{fn:5}, p.~\pageref{fn:5}. There we discussed
the problem that there is a freedom in relating a two-vertex to
Neumann boundary state on a circle. Indeed,
instead of $f_1(z)={z}$ and $f_2(z)= \frac1z$ we could use
general \gl2 transformation thereof. Then we would get
\begin{equation}
  \label{eq:131}
    h_1(z) = \frac{w(z)\,a + b}{w(z)\, c + d},
    \quad
    h_2(z) = \frac{\frac{a}{\bw(z)} + b}{\frac c{\bw(z)}+ d}
\end{equation}
instead of~(\ref{eq:130}). Or inverse
\begin{equation}
  \label{eq:132}
  w(z) = \left(-\frac dc\right) \frac{ h_1(z) - b/d}{ h_1(z) - a/c}, \quad
  {\bar w}(z) = \left(-\frac cd\right) \frac{ h_2(z) - a/c}{ h_2(z) - b/d}
\end{equation}
Up until now, we did not specify
analytic properties of $h_I(z)$. It turns out that by comparing them with
analytic properties of $w(z)$ one can fix this \gl2 freedom.
For the purpose of identification between Toda tau-function and Neumann
boundary state on a curve we considered only univalent $w$ and
$\bw$~(\ref{eq:6}), i.e.
\begin{equation}
  \label{eq:133}
  w(z) = \frac zr + p_0 +\frac{p_1}z +  \CO(\frac1{z^2})
\end{equation}
Now eq.~(\ref{eq:130}) assumes that $h_1$ is univalent too but
expansion of $h_2$ starts with $1/z$.
On the other hand in future applications $h_1$ and $h_2$ are 
taken to be regular at infinity
\begin{equation}
  \label{eq:134}
  h_I(z) = h_I(\infty) - \frac{[h'_I](\infty)}z + \CO(\frac1{z^2}),\;I=1,2
\end{equation}
Insisting on both of the conditions~(\ref{eq:133}),~(\ref{eq:134}) as well as
on eq.~(\ref{eq:132}) will fix coefficients of \gl2 transformation. It is easy
to see that the first of the equations in~(\ref{eq:132})
requires\footnote{
  We do not consider the case when $h_1(\infty)=h_2(\infty)$.}
$a/c = h_1(\infty)$ and the second gives $b/d = h_2(\infty)$.  
The scaling coefficient $c/d$ is left unfixed for the moment. 
We can calculate how $r$ and $\bar r$ in the expansion
of $w$ and $\bw$ depend on corresponding coefficients in $h$.
It is
\begin{equation}
  \frac 1r = \left(- \frac dc \right) \frac{h_1(\infty) - h_2(\infty)}{ [h'_1](\infty)},
  \quad
  \frac 1{\bar r} = \left(-\frac cd \right) \frac{h_2(\infty) - h_1(\infty)}{ [h'_2](\infty)}
\end{equation}
If we require ${\bar r}$ to be complex conjugate of $r$ then
the factor $d/c$ is determined. For the case when all $h_I(\infty)$
and $[h'_I](\infty)$
are real we have
\begin{equation}
\frac dc = - \sqrt{- \frac{[h'_1](\infty)}{[h'_2](\infty)}}
\end{equation}
Let us give for completeness the final result
\begin{eqnarray}
  \label{eq:res74}
 &&w(z) = \left(-\frac dc\right) \frac{ h_1(z) - b/d}{ h_1(z) - a/c}, \quad
  {\bar w}(z) = \left(-\frac cd\right) \frac{ h_2(z) - a/c}{ h_2(z) - b/d} \nonumber\\
 &&\frac ac = h_1(\infty)~~~~~~~~
  \frac bd = h_2(\infty)~~~~~~
  \frac dc = - \sqrt{- \frac{[h'_1](\infty)}{[h'_2](\infty)}}
\end{eqnarray}

\section{The identification between dToda and a two-vertex}
\label{app:2vert-toda}

In this Appendix we check explicitly if identification between
Neumann coefficients of two-vertex and second derivatives
of Toda tau-function $F$ is consistent. The formulation of the problem
is the following. The second derivatives of $F$ as second derivatives 
of any Toda tau-function satisfy Hirota equations. We want to plug
Neumann coefficients in the corresponding equations and see that
they are still satisfied.
For convenience let us list Hirota equations, definitions of Neumann coefficients,
and the identification between them and second derivatives of $F$.
Hirota equations are
\begin{eqnarray}
\partial^2_{t_0}F &=& \log |r|^2 \label{eq:1.1}\\
\partial_{t_0} \partial_{t_n} F &=&
\oint \frac{dz}{2\pi i} z^{n} 
\partial_z \log{\LB w(z) r \RB}~~~{\rm for}~~n>0 \label{eq:1.2} \\
\partial_{t_n}\partial_{t_k} F &=& \oint \oint \frac{dzd\zeta}{(2\pi i)^2}
z^{n} \zeta^{k}
\partial_z \partial_\zeta \log{\frac{w(z) - w(\zeta)}{z-\zeta}} \label{eq:1.3}\\
\partial_{t_n}\partial_{{\bar t}_k} F &=& - \oint \oint \frac{dzd\zeta}{(2\pi i)^2}
z^{n} \zeta^{k}
\partial_z \partial_\zeta \log \LB 1 - \frac{1}{w(z){\bar w}(\zeta)}
\frac{e^{\partial^2_{t_0}F}}{|r|^2} \RB 
\label{eq:1.4}
\end{eqnarray}
Equations (\ref{eq:1.2}) completely determine the conformal 
transformation $w(z)$ and plugging it in eq.~(\ref{eq:1.3}) and  
eq.~(\ref{eq:1.4})
one can
calculate the rest of the derivatives of $F$. The analytic properties
of $w(z)$ are
\bea
w(z) = \frac z r  + \sum_{k=0}^\infty \frac{p_k}{z^k}
\eea
Now we repeat definition of Neumann 
coefficients. They depend on conformal maps $h_1,h_2$ as
\begin{eqnarray}
\label{eq:3.1}
N_{km}^{IJ} &=& \frac{1}{km}\oint \frac{dz}{2\pi i} \oint \frac{d\zeta}{2\pi i}
z^k \zeta^m \de_z \de_\zeta\log\Big( h_I(z) - h_J(\zeta) \Big)\\
\label{eq:3.2}
N_{0m}^{IJ} &=& - \frac{1}{m}\oint \frac{dz}{2\pi i} z^m \de_z \log\Big(
h_I(\infty)-h_J(z)\Big)\\ 
\label{eq:3.3}
  N_{00}^{IJ}&=&\left\{
  \begin{array}[c]{ll}
    \log ([h'_I](\infty))& I=J\\
    \log (h_I(\infty)-h_J(\infty))& I\neq J
  \end{array}\right.
\end{eqnarray}
where $\displaystyle [h'](z) \equiv -\frac{d h(z)}{d 1/z}$.
And finally we give the full list of identifications
\begin{eqnarray}
\label{eq:2.3}
\p_{\t0}^2 F& =& \hphantom{-}N_{00}^{11} + N_{00}^{22} - N_{00}^{12} - N_{00}^{21}\\
\frac{1}{k}\partial_{t_0}\partial_{t_k} F& =& \hphantom{-} N^{11}_{0k} - N^{21}_{0k} 
\label{eq:2.2}\\
\frac{1}{k}
\partial_{t_0}\partial_{{\bar t}_k} F& = &\hphantom{-} N^{22}_{0k} - N^{12}_{0k}
\label{eq:2.1}\\
\frac{1}{nk}\partial_{t_n}\partial_{t_k} F& =&  \hphantom{-} N^{11}_{nk} \label{eq:2.6}\\
\frac{1}{nk}
\partial_{{\bar t}_n}\partial_{{\bar t}_k} F& = &  \hphantom{-}N^{22}_{nk}
\label{eq:2.4}\\
\frac{1}{nk}\partial_{t_n}\partial_{{\bar t}_k} F& =&-N^{12}_{nk}
\label{eq:2.5} 
\end{eqnarray}
Our strategy is the following now. From identifications in zero-mode sector,
eqs.~(\ref{eq:2.3})-(\ref{eq:2.1}) 
we find the relation between conformal maps $h$ and $w$.
\begin{eqnarray}
&& |r|^2 = - \frac{[h'_1](\infty)[h'_2](\infty)}{\LB h_1(\infty)-h_2(\infty) \RB^2}\\
&& w(z) = -\frac{d h_1(z) - b}{c h_1(z) - a} 
\label{eq:135}\\
&& {\bar w}(z) = 
- \frac{c h_2(z) - a}{d h_2(z) - b} 
\label{eq:136}
\end{eqnarray}
where coefficients $a,b,c,d$ are the same as in eq.~(\ref{eq:res74}).
They were obtain there by relating the two-vertex to Neumann
boundary state and requiring both $h(z)$ to be regular at infinity.

If we plug this in place of $w$ in eq.~(\ref{eq:1.4}) for  
$\partial_{t_k} \partial_{{\bar t}_n} F$ we should get corresponding Neumann
coefficients $N^{12}_{kn}$ as predicted in  eq.~(\ref{eq:2.5})
That is we should prove that
\begin{eqnarray}
\partial_{t_n}\partial_{{\bar t}_k} F &=& -
\oint \oint \frac{dzd\zeta}{(2\pi i)^2}
z^{n} \zeta^{k} \partial_z \partial_\zeta 
\log{\Big(1 - \frac{1}{w(h_1(z)){\bar w}(h_2(\zeta))}
\frac{e^{\partial^2_{t_0}F}}{r{\bar r}}\Big)} \nonumber \\
&=& -
\oint \frac{dz}{2\pi i} z^{n} \zeta^{k}
\partial_z \partial_\zeta \log{\Big( h_1(z) - h_2(\zeta) \Big)} \nonumber \\ 
&=& -
nk N^{12}_{nk}
\end{eqnarray}
Let us plug maps $w$ and $\bar w$ found in eq.~(\ref{eq:135})-eq.~(\ref{eq:136}),
into the above equation.
\begin{eqnarray}
\partial_{t_n}\partial_{{\bar t}_k} F &=& \nonumber \\
&-&\oint \oint \frac{dzd\zeta}{(2\pi i)^2}
z^{n} \zeta^{k}
\partial_z \partial_\zeta \log{\LB 1 -
\frac{h_1(z) - h_2(\infty)}{h_1(z) - h_1(\infty)} 
\frac{h_2(\zeta) - h_1(\infty)}{h_2(\zeta) - h_2(\infty)}
\RB} = \nonumber \\
&-& \oint \oint \frac{dzd\zeta}{(2\pi i)^2}
z^{n} \zeta^{k}
\partial_z \partial_\zeta \log{\Bigg( h_2(\zeta)- h_1(z)
\Bigg)}
\end{eqnarray}
It is indeed equal to $- nk N^{12}_{nk}$.  Similar checks can be performed for
identifications in eq.~(\ref{eq:2.6}) and eq.~(\ref{eq:2.4}). The checks are
satisfied.

\end{document}